\documentclass[
 reprint,
 superscriptaddress,
 amsmath,amssymb,
 aps,
 prb,
 floatfix,
]{revtex4-2}

\usepackage{graphicx}
\usepackage{dcolumn}
\usepackage{bm}
\usepackage{braket}
\usepackage{physics}
\usepackage{ulem}
\usepackage[unicode=true]{hyperref}
\usepackage{xcolor}

\newcommand{\hc}{\mathrm{h.c.}}
\usepackage[T1]{fontenc}

\definecolor{ori_b}{HTML}{001c7f}
\definecolor{ori_b1}{HTML}{0025a8}
\definecolor{ori_b2}{HTML}{0f36bd}
\definecolor{ori_y}{HTML}{d5bb67}
\definecolor{ori_g}{HTML}{12711c}
\definecolor{ori_r}{HTML}{8c0800}
\hypersetup{
  colorlinks = true,       
  linkcolor  = ori_b2, 
  filecolor  = ori_b1, 
  urlcolor   = ori_b, 
  citecolor  = ori_b1  
}
\begin{document}

\title{Lattice-geometry dependence of Type-I and Type-II magnetic flux responses in strongly correlated bosons}

\author{Takayuki Yokoyama}
\email[]{yokotaka308@hiroshima-u.ac.jp}
\affiliation{Quantum Matter Program, Graduate School of Advanced Science and Engineering, Hiroshima University, Higashihiroshima, Hiroshima 739-8530, Japan}

\author{Yasuhiro Tada}

\affiliation{Quantum Matter Program, Graduate School of Advanced Science and Engineering, Hiroshima University,
Higashihiroshima, Hiroshima 739-8530, Japan}
\affiliation{Institute for Solid State Physics, University of Tokyo, Kashiwa 277-8581, Japan}

\begin{abstract}
Superfluids and superconductors are classified into Type-I and Type-II according to their response to a magnetic flux, but this classification has hardly been addressed for lattice bosons under magnetic flux, where the Meissner and vortex phases have been studied mostly in ladder geometries.
Here, we study the flux response of the extended Bose-Hubbard model with nearest-neighbor interaction $V$ at half filling on square and triangular ladders with two to four legs by using the density-matrix renormalization group.
On the square ladders, the vortex phase is confined to a region near $V=0$. At finite $V$, a magnetic flux-induced charge-density-wave (CDW) order replaces it, and the response becomes Type-I-like. In contrast, on the triangular ladders, the vortex phase persists over the whole range of $V$ studied, and the response remains Type-II-like.
By extrapolating the ladder results to two dimensions, we argue that the Type-I or Type-II character of the flux response depends on the lattice geometry, which suggests a role of geometric frustration.
\end{abstract}
\maketitle

\section{\label{sec:introduction}Introduction}

Superfluids and superconductors, understood as consequences of the spontaneous breaking of a global U(1) symmetry, have been a central topic of quantum physics.
An important feature of these states is their response to a magnetic field: the Meissner effect, in which the field is expelled, and the formation of vortices, in which the field penetrates the bulk.
Superconductors are conventionally classified into Type-I and Type-II according to their response to an applied magnetic field. In Type-I superconductors, the Meissner state is destroyed at a critical magnetic field, whereas Type-II superconductors exhibit an intermediate mixed state in which magnetic flux penetrates in the form of vortices.
In solid state physics, this distinction is determined by the ratio of the magnetic penetration depth to the coherence length, and the vortex states in strong magnetic fields have been extensively discussed~\cite{Abrikosov2004, Tinkham2004}.

Superfluids under a magnetic flux have also been investigated in cold-atom systems.
In optical lattices~\cite{Gross2017,Schafer2020}, artificial gauge fields generated by laser-assisted tunneling allow us to study the flux response of neutral atoms~\cite{Aidelsburger2013,Miyake2013,Aidelsburger2018}.
More recently, the effects of artificial gauge fields have also been discussed in Rydberg atom arrays ~\cite{Browaeys2020,Morgado2021,Lienhard2020,Wu2022,Chen2024,Eix2025}.
In particular, bosons on ladders with two or three legs show a Meissner phase and a vortex phase as the flux increases, which have been discussed theoretically by field-theoretical approaches such as bosonization~\cite{Giamarchi2003,Orignac2001,Tokuno2014,Uchino2015,Uchino2016,Citro2018} and numerical calculations~\cite{Dhar2012,Dhar2013, Petrescu2013,Piraud2015,Dio2015,Greschner2015,Kolley2015,Natu2015,Takayuki2026}, and have also been observed experimentally~\cite{Atala2014,Stuhl2015,Impertro2025}.
Triangular ladders have also been studied recently~\cite{Halati2023,Li2023,Baldelli2024,Zhang2026,Dasgupta2026,Jian2026}.

For strongly correlated bosons under a magnetic flux in cold-atom systems, previous studies show that the response to the magnetic flux is almost always Type-II-like, in contrast to solids, in which there are both Type-I and Type-II superconductors~\cite{Tinkham2004}.
The Meissner phase is followed by a vortex phase as the flux is increased in the previous studies~\cite{Piraud2015,Kolley2015,Halati2023,Atala2014,Impertro2025}, while a Type-I-like response, in which the Meissner phase is directly replaced with a non-vortex phase without the intervening vortex phase, has hardly been reported.
It is not known why the Type-II-like response dominates, nor where a Type-I-like response can be found.
The magnetic flux response of bosonic ladders has not been systematically investigated from this viewpoint, and what controls the type remains an open question.

In this work, we investigate how interactions and lattice geometry affect the magnetic flux response of strongly correlated bosons.
We study hard-core bosons at half filling with a nearest-neighbor repulsion $V$ on square and triangular ladders with two to four legs by using the density-matrix renormalization group (DMRG)~\cite{White1992,SCHOLLWOCK2011,McCulloch2007}.
We classify the magnetic flux response as Type-I-like or Type-II-like according to whether the vortex phase appears between the Meissner phase and the non-vortex phase.
Our results are summarized in Fig.~\ref{fig:sketch}. On the square ladders, we find that the vortex phase is confined to a region near $V=0$ and is replaced with a flux-induced charge-density-wave (CDW) order at finite $V$, so that the response becomes Type-I-like.
On the triangular ladders, the vortex phase is robust against $V$, and the response remains Type-II-like in contrast to the square ladders.
Extrapolating the ladder results to two-dimensional systems, we suggest that the Type-I or Type-II character of the magnetic flux response depends on the lattice geometry, in particular on the geometric frustration.
This viewpoint goes beyond the Meissner and vortex states of cold-atom ladders, and provides a guide to the flux response of superfluids, superconductors, and quantum magnets on general lattices.

\section{\label{sec:model}Details of numerical calculation}
\begin{figure}[t]
  \centering
  \includegraphics[width=\columnwidth]{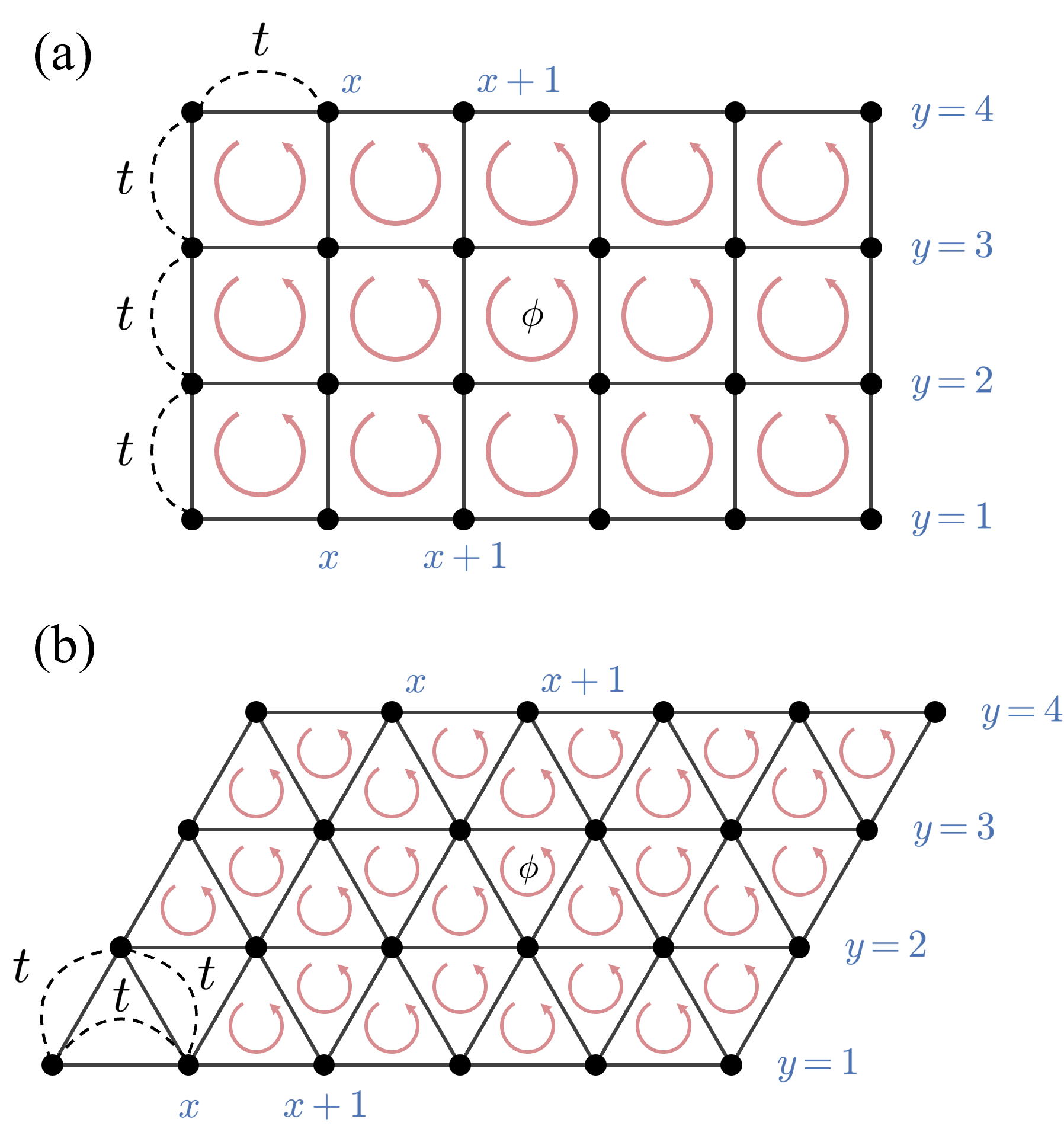}
  \caption{Lattice geometries of (a) the square ladder and (b) the triangular ladder with $N_{\mathrm{leg}}=4$.
  Sites are labeled by $(x,y)$ with $x=1,\dots,L$ along the legs and $y=1,\dots,N_{\mathrm{leg}}$.
  Dashed lines indicate examples of the hopping $t$, and circular arrows indicate the uniform flux $\phi$ threading each plaquette.}
  \label{fig:lattice}
\end{figure}

\subsection{Extended Bose-Hubbard model under a magnetic flux}
\label{sec:model_explain}
We study the extended Bose-Hubbard model on square and triangular ladders under a uniform magnetic flux $\phi$ (Fig.~\ref{fig:lattice}).
The Hamiltonian is
\begin{align}
H = &- t \sum_{\langle i,j\rangle}
\left( e^{iA_{ij}} b_i^\dagger b_j + \hc \right) \nonumber \\
&+ \frac{U}{2} \sum_i n_i (n_i-1)
+ V \sum_{\langle i,j\rangle}
\left(n_i-\frac{1}{2}\right)
\left(n_j-\frac{1}{2}\right),
\label{eq:HCB}
\end{align}
where $b_i^\dagger$ ($b_i$) creates (annihilates) a boson on site $i$, $n_i=b_i^\dagger b_i$, and $A_{ij}$ is the Peierls phase that describes the magnetic flux.
The site is labeled by $i=(x_i,y_i)$ with $x_i=1,\cdots,L$ and $y_i=1,\cdots,N_{\rm leg}$, where $L$ is the system length and $N_{\rm leg}$ is the number of legs.
The bonds $\langle i,j\rangle$ are the nearest-neighbor bonds along the legs and rungs; in the triangular ladder, they also include the diagonal bonds.
The hopping amplitude is $t>0$, $U$ is the on-site repulsion, and $V$ is the nearest-neighbor repulsive interaction.
This setup is motivated mainly by ultracold atoms in optical lattices, where the Peierls phase can be engineered as an artificial gauge field by laser-assisted tunneling \cite{Aidelsburger2013,Miyake2013,Aidelsburger2018,Browaeys2020,Morgado2021,Lienhard2020,Wu2022,Chen2024}.
Bosonic ladders under such a gauge flux have been realized experimentally, and the Meissner and vortex states have been observed, including triangular ladders \cite{Atala2014,Stuhl2015,Impertro2025,Li2023}.
The nearest-neighbor interaction $V$ can arise, for example, from dipole--dipole interactions in optical lattices~\cite{Su2023,Pascal2025} or from interactions between Rydberg atoms in optical tweezer arrays~\cite{Browaeys2020,Morgado2021,Bernien2017}.

To clarify the flux response in the simplest setup, we focus on the hard-core limit ($U\to\infty$) at half filling.
In this limit, through $b_i^\dagger\to S_i^+$ and $n_i-1/2\to S_i^z$, the model maps onto the spin-$1/2$ XXZ model,
\begin{equation}
H=-\sum_{\langle i,j\rangle}\left(\frac{J_{xy}}{2}\,e^{iA_{ij}}S_i^{+}S_j^{-}+\hc\right)+J_z\sum_{\langle i,j\rangle}S_i^{z}S_j^{z},
\label{eq:XXZ}
\end{equation}
with $J_{xy}=2t$ and $J_z=V$, so that $V/2t=J_z/J_{xy}$ is the Ising anisotropy.
$V/2t=1$ corresponds to $J_z/J_{xy}=1$.
This mapping helps us understand the effects of the flux, depending on whether the in-plane interaction is ferromagnetic or antiferromagnetic. We will refer to this point when we discuss distinct symmetries of the Hamiltonians for square and triangular ladders associated with a gauge transformation.

We use the Landau gauge
\begin{equation}
A_{ij} =
\begin{cases}
    \phi\,\dfrac{x_i+x_j}{2}\,(y_i-y_j) & \text{(square)}, \\[2mm]
    2\phi\,\dfrac{x_i+x_j}{2}\,(y_i-y_j) & \text{(triangular)},
\end{cases}
\label{eq:gauge}
\end{equation}
where $\phi$ is the magnetic flux per plaquette (each square or triangle). 
We note the symmetries of the model.
The complex conjugation of the Hamiltonian changes the sign of the Peierls phase, $A_{ij}\to-A_{ij}$, so that the spectrum is invariant under $\phi\to-\phi$.
In addition, the flux is defined modulo the flux quantum, $\phi\to\phi+2\pi$, by a local U(1) gauge transformation.
These two relations lead to $\phi\to2\pi-\phi$, so that it is sufficient to consider $0\le\phi\le\pi$, as we do in this work.

In this study, we mainly calculate the ground states of two-, three- and four-leg ladders.
We use the DMRG method~\cite{White1992,SCHOLLWOCK2011,McCulloch2007} with ITensor.jl~\cite{ITensor}.
We fix the system size $L=60$ under the open boundary condition, which is comparable to the sizes used in previous DMRG studies of bosonic flux ladders~\cite{Piraud2015,Kolley2015,Halati2023}. The DMRG calculations conserve the total particle number (U(1) symmetry) and use a maximum bond dimension of $\chi=800$.
The truncation error is typically smaller than $10^{-5}$, which is sufficiently small for determining the qualitative features of the phase diagrams.

\subsection{Gauge transformation and sign of hopping $t$}
\label{sec:gauge}
We consider the gauge transformation $b_j\to(-1)^{x_j+y_j}b_j$, which does not change the Peierls phase $A_{ij}$ and the interaction terms.
It changes the sign of the hopping on a bond $(i,j)$ when the parities of $x_i+y_i$ and $x_j+y_j$ are different.

On the square lattice, every bond connects sites with different parities, so that the sign of the hopping is changed on every bond for any flux,
\begin{equation}
H(t,\phi) \equiv  H(-t,\phi)\qquad\text{for any }\phi,
\label{eq:sq_sign}
\end{equation}
where $\equiv $ denotes gauge equivalence.
Namely, the sign of $t$ is irrelevant on the square lattice, and in the spin language the ferromagnetic and the antiferromagnetic in-plane interactions are equivalent at any flux.

In contrast, the $\pi$-flux has a further meaning for the triangular lattice.
At $\pi$-flux ($\phi=\pi$) in Eq.~\eqref{eq:gauge}, the Peierls phase is $0$ on the leg and rung bonds and $\pi$ on the diagonal bonds.
The same gauge transformation changes the sign of the hopping on the leg and rung bonds but not on the diagonal bonds, since the diagonal bond connects sites with the same parity, and we obtain
\begin{equation}
H(t,\phi=\pi) \equiv  H(-t,\phi=0).
\label{eq:tri_sign}
\end{equation}
In the spin language, the ferromagnetic in-plane interaction at $\phi=0$ is turned into an antiferromagnetic one at $\phi=\pi$.

\subsection{Observables}
\label{sec:Observables}
We examine the Meissner and vortex states based on the local currents.
The boson current on the bond $(i,j)$ is defined as $j_{ij}\equiv\partial H/\partial A_{ij}$,
\begin{equation}
j_{ij}=-it\left(e^{iA_{ij}}\,b_i^\dagger b^{}_j-\hc\right),
\label{eq:jij}
\end{equation}
which is the current flowing from the site $i$ to the site $j$.
We distinguish the currents on the leg bonds, the rung bonds, and, in the triangular ladder, the diagonal bonds.
In the Landau gauge [Eq.~\eqref{eq:gauge}], the leg bonds carry no Peierls phase, and the leg current flowing from $(x+1,y)$ to $(x,y)$ is
\begin{equation}
j^{\parallel}_{x,y}=-it\left(b^\dagger_{x+1,y}\,b^{}_{x,y}-\hc\right).
\label{eq:jleg}
\end{equation}
The rung bond connecting $(x,y)$ and $(x,y+1)$ carries the Peierls phase
\begin{equation}
A_x=
\begin{cases}
\phi\,x & \text{(square)},\\
2\phi\,x & \text{(triangular)},
\end{cases}
\label{eq:Ax}
\end{equation}
and the rung current flowing from the leg $y+1$ to the leg $y$ is
\begin{equation}
j^{\perp}_{x,y}=-it\left(e^{iA_x}\,b^\dagger_{x,y+1}\,b^{}_{x,y}-\hc\right).
\label{eq:jperp}
\end{equation}
In the triangular ladder, there is an additional diagonal bond connecting $(x,y+1)$ and $(x+1,y)$, which we label by its midpoint $x+1/2$.
Its Peierls phase is $A_{x+1/2}=2\phi\,(x+1/2)$, i.e., Eq.~\eqref{eq:Ax} evaluated at the midpoint, and the diagonal current flowing from the leg $y+1$ to the leg $y$ is
\begin{equation}
j^{\times}_{x+\frac12,y}=-it\left(e^{iA_{x+\frac12}}\,b^\dagger_{x,y+1}\,b^{}_{x+1,y}-\hc\right),
\label{eq:jdiag}
\end{equation}
where $j^{\times}\equiv0$ in the square ladder.
Although the operators are written in this gauge, the expectation values of the currents are gauge invariant.

To suppress edge effects with open boundary conditions, all observables below are averaged over the middle region $M_x=\{x_l,x_l+1,\dots,x_r\}$ with $x_l=L/4+1$ and $x_r=3L/4$, which has $N_x=L/2$ columns and $N_s=N_xN_{\rm leg}$ sites.
Only the bonds whose two ends are inside $M_x$ are included in the sums.
The rung bonds lie in a single column, so that their sums run over $x=x_l,\dots,x_r$, whereas the leg and the diagonal bonds connect the columns $x$ and $x+1$, so that their sums run over $x=x_l,\dots,x_r-1$, which is one fewer than the number of columns.
Accordingly, the numbers of the bonds in the sums are $N^{\parallel}=(N_x-1)N_{\rm leg}$ for the leg bonds, $N^{\perp}=N_x(N_{\rm leg}-1)$ for the rung bonds, and $N^{\times}=(N_x-1)(N_{\rm leg}-1)$ for the diagonal bonds, where the diagonal bonds exist only in the triangular ladder.

In the Meissner state, the currents flowing along the two edge legs $y=1$ and $y=N_{\rm leg}$ have opposite directions.
We therefore define the chiral current as the order parameter of the Meissner phase by the difference of the two edge currents,
\begin{equation}
\mathcal{O}_{\rm M}=\frac{1}{2(N_x-1)}\sum_{x=x_l}^{x_r-1}\ev{j^{\parallel}_{x,1}-j^{\parallel}_{x,N_{\rm leg}}},
\label{eq:Jc}
\end{equation}
where $\mathcal{O}_{\rm M}>0$ corresponds to the diamagnetic response.

In the vortex phase, finite transverse currents appear when vortices penetrate the ladder.
Therefore, we use the averaged transverse-current amplitude as an order parameter of the vortex phase,
\begin{align}
|\mathcal{J}_{\rm rung}|=\;&\frac{1}{N^{\perp}}\sum_{y=1}^{N_{\rm leg}-1}\sum_{x=x_l}^{x_r}\abs{\ev{j^{\perp}_{x,y}}} \nonumber\\
&+\begin{cases}
0 & \text{(square)},\\[2mm]
\dfrac{1}{N^{\times}}\displaystyle\sum_{y=1}^{N_{\rm leg}-1}\sum_{x=x_l}^{x_r-1}\abs{\ev{j^{\times}_{x+\frac12,y}}} & \text{(triangular)}.
\end{cases}
\label{eq:Jr}
\end{align}
The absolute values are taken to avoid cancellation between currents flowing in opposite directions in the vortex phase.

To characterize the vortex pattern, we introduce the Fourier transform of the
currents along the leg direction,
\begin{align}
  j^{\perp}_y(k) &= \frac{1}{\sqrt{N_x}} \sum_{x=x_l}^{x_r} e^{-ikx}
     \Bigl( \bigl\langle j^{\perp}_{x,y} \bigr\rangle - \overline{j^{\perp}_y} \Bigr), \label{eq:jperpk}\\
  j^{\times}_y(k) &= \frac{1}{\sqrt{N_x-1}} \sum_{x=x_l}^{x_r-1} e^{-ik(x+\frac{1}{2})}
     \Bigl( \bigl\langle j^{\times}_{x+\frac12,y} \bigr\rangle - \overline{j^{\times}_y} \Bigr), \label{eq:jdiagk}
\end{align}
where
\begin{equation}
  \overline{j^{\perp}_y} = \frac{1}{N_x} \sum_{x=x_l}^{x_r} \bigl\langle j^{\perp}_{x,y} \bigr\rangle, \qquad
  \overline{j^{\times}_y} = \frac{1}{N_x-1} \sum_{x=x_l}^{x_r-1} \bigl\langle j^{\times}_{x+\frac12,y} \bigr\rangle
  \label{eq:javg}
\end{equation}
are the averages over the middle region $M_x$.
The averages are subtracted because, in the triangular ladders with three or more legs,
static transverse currents flow even in the Meissner phase; they do not form a vortex
pattern but would otherwise give the largest weight at $k \to 0$.
The corresponding rung-current structure factor is
\begin{equation}
  S_{\rm rung}(k) = \frac{1}{N_{\rm leg}-1} \sum_{y=1}^{N_{\rm leg}-1}
  \Bigl( \bigl| j^{\perp}_y(k) \bigr|^2 + \bigl| j^{\times}_y(k) \bigr|^2 \Bigr).
  \label{eq:Srung}
\end{equation}
A peak of $S_{\rm rung}(k)$ at $k = k_{\rm v}$ signals a vortex pattern with the period
$2\pi/k_{\rm v}$ along the ladder. We define the vortex density, i.e., the number of
vortices per lattice spacing along the legs, from the position of the largest peak of
$S_{\rm rung}(k)$,
\begin{equation}
  \rho_{\rm v} = \frac{k_{\rm v}}{2\pi}, \qquad
  k_{\rm v} = \mathop{\arg\max}_{0<k\le\pi} S_{\rm rung}(k),
  \label{eq:rhov}
\end{equation}
and we set $\rho_{\rm v} = 0$ when $\max_{0<k\le\pi} S_{\rm rung}(k)$ is below the threshold
$\delta_{\rm v} = 10^{-3}$, which corresponds to the Meissner phase, where the rung and
diagonal currents show no periodic modulation. The maximal value $\rho_{\rm v} = 1/2$
($k_{\rm v} = \pi$) corresponds to a vortex pattern with a period of two lattice spacings.
We note that a commensurate vortex phase with $\rho_{\rm v}=1/2$ ($k_{\rm v}=\pi$) is distinguished from the state with $\rho_{\rm v}=0$ by the peak of $S_{\rm rung}(k)$.
In contrast, uniform rung currents ($k_{\rm v}=0$), corresponding to a commensurate vortex phase with a period of one lattice spacing, give $\rho_{\rm v}=0$ under our definition and cannot be distinguished from the Meissner phase by $\rho_{\rm v}$ alone~\footnote{Since our focus is not on distinguishing different vortex structures, such as commensurate and incommensurate vortex phases, we adopt this simplified definition for the vortex density.}.
We therefore use $\rho_{\rm v}$ together with $|\mathcal{J}_{\rm rung}|$.

For the square ladders, we also define the staggered current along the legs,
\begin{equation}
\mathcal{J}_{\rm sc}=\frac{1}{N^{\parallel}} \sum_{y=1}^{N_{\rm leg}} \sum_{x=x_l}^{x_r-1}(-1)^{x+y}\ev{j^{\parallel}_{x,y}},
\label{eq:Jsc}
\end{equation}
which is the order parameter of the staggered-current state, in which the circulating currents alternate their chirality on neighboring plaquettes.
The sign of $\mathcal{J}_{\rm sc}$ is chosen spontaneously in the DMRG calculations.

The nearest-neighbor repulsion $V$ is naively expected to induce a CDW order, in which the particle density is staggered between neighboring sites.
We therefore analyze the CDW order parameter
\begin{equation}
\mathcal{O}_{\rm CDW}=\frac{1}{N_s} \sum_{y=1}^{N_{\rm leg}} \sum_{x=x_l}^{x_r} (-1)^{x+y}\,\ev{n_{x,y}},
\label{eq:CDW}
\end{equation}
where the sign $(-1)^{x+y}$ corresponds to the wave vector $\bm{Q}$ satisfying $\bm{Q}\cdot\bm{a}_1=\bm{Q}\cdot\bm{a}_2=\pi$ for the primitive vectors $\bm{a}_1=\bm{e}_x$ and $\bm{a}_2=\bm{e}_y$ (square) or $\bm{a}_2=\bm{e}_x/2+\sqrt{3}\bm{e}_y/2$ (triangular), where $(\bm{e}_x,\bm{e}_y)$ is an orthogonal basis.
It is the wave vector $(\pi,\pi)$ of the CDW order for the square lattice, and the $M$ point for the triangular lattice, where the density is staggered along the legs and the rungs but uniform along the diagonal bonds.

For the bond order on the triangular ladder, we define the bond energies on the rung bond and on the diagonal bond as
\begin{align}
\epsilon^{\perp}_{x,y}&=-t\ev{e^{iA_x}\,b^\dagger_{x,y+1}\,b^{}_{x,y}+\hc},\\
\epsilon^{\times}_{x+\frac12,y}&=-t\ev{e^{iA_{x+\frac12}}\,b^\dagger_{x,y+1}\,b^{}_{x+1,y}+\hc},
\label{eq:eps}
\end{align}
which are the real parts of the same expectation values whose imaginary parts give the currents $j^{\perp}_{x,y}$ and $j^{\times}_{x+\frac12,y}$.
The rung bonds and the diagonal bonds form zigzag chains between the neighboring legs, and we define the order parameter for the bond order on the zigzag chains, in which the rung and the diagonal bonds are alternately strong and weak, as
\begin{align}
\mathcal{O}_{\rm bond} = \frac{1}{N^{\times}}\sum_{y=1}^{N_{\rm leg}-1}\sum_{x=x_l}^{x_r-1}\epsilon^{\times}_{x+\frac12,y} - \frac{1}{N^{\perp}}\sum_{y=1}^{N_{\rm leg}-1}\sum_{x=x_l}^{x_r}\epsilon^{\perp}_{x,y}.
\label{eq:bond}
\end{align}
Here the sign is $-$ for the rung bonds and $+$ for the diagonal bonds, so that $\mathcal{O}_{\rm bond}$ is finite when the rung and the diagonal bonds are not equivalent.

\begin{figure*}[t]
  \centering
  \includegraphics[width=\textwidth]{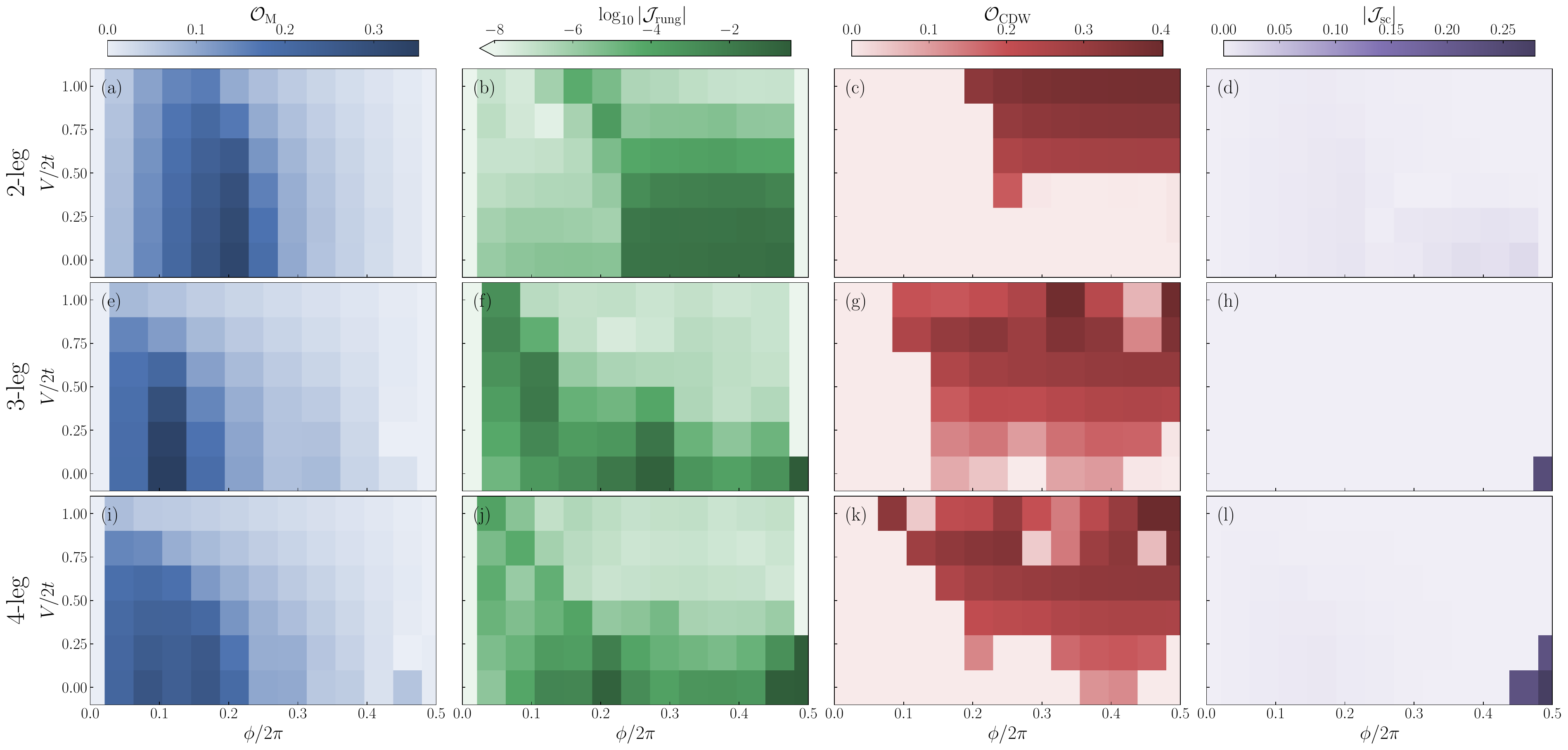}
  \caption{Magnetic flux response of (top) two-,
(middle) three-, and (bottom) four-leg
square ladders for $L=60$.
  From left to right, the panels show the diamagnetic current $\mathcal{O}_{\mathrm{M}}$, the logarithm of the rung-current magnitude $\log_{10}|\mathcal{J}_{\mathrm{rung}}|$, the CDW order parameter $\mathcal{O}_{\mathrm{CDW}}$, and the staggered current $|\mathcal{J}_{\mathrm{sc}}|$, as functions of the flux $\phi / 2 \pi$ and the nearest-neighbor interaction $V/2t$.}
  \label{fig:flux_response}
\end{figure*}

\section{Magnetic flux response in square ladders}
\label{sec:square}

\begin{figure*}[t]
  \centering
  \includegraphics[width=\textwidth]{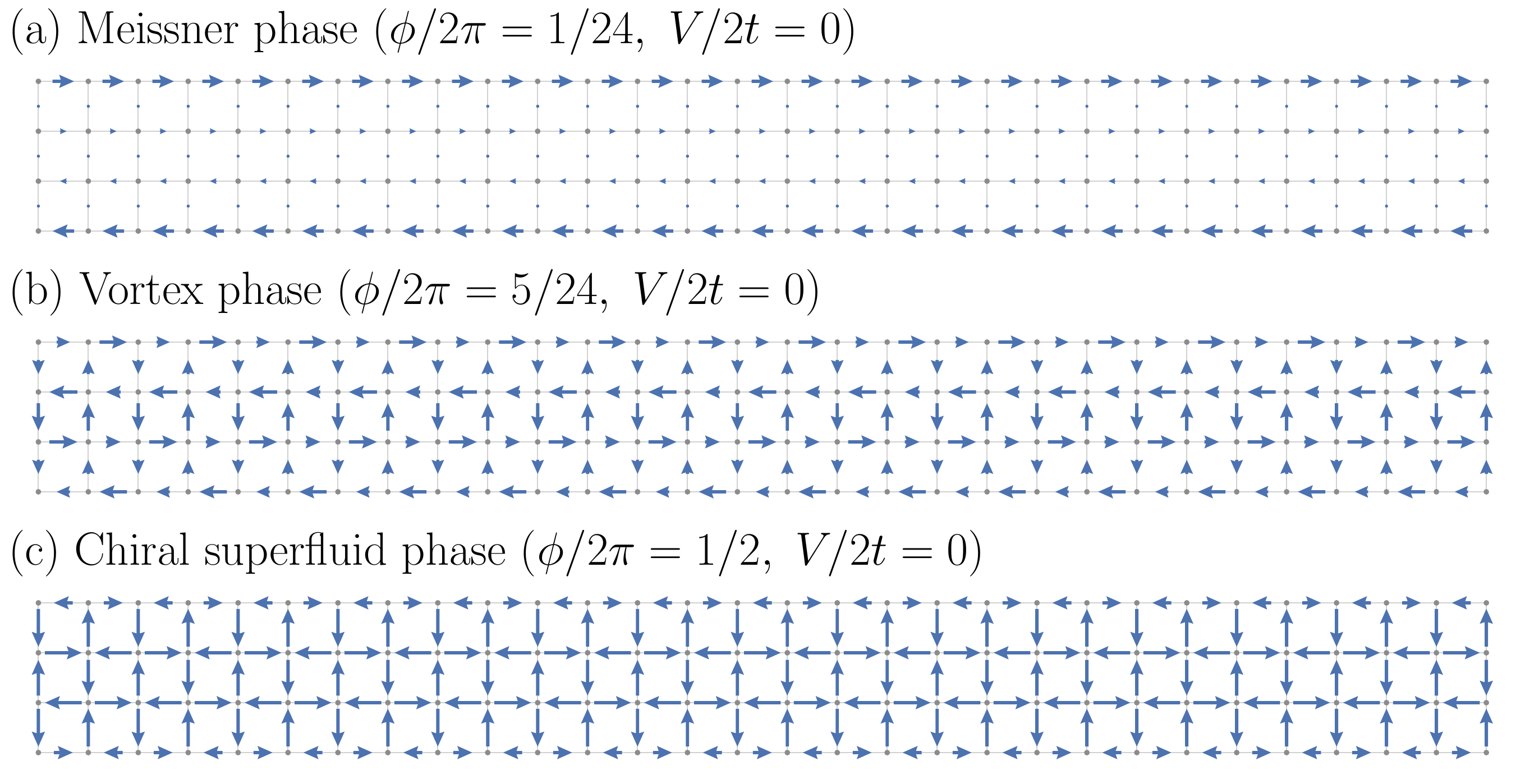}
  \caption{Real-space patterns of the bond currents in the four-leg square ladder at $V/2t=0$ for (a) the Meissner phase ($\phi/2\pi=1/24$), (b) the vortex phase ($\phi/2\pi=5/24$), and (c) the chiral superfluid (staggered current) phase ($\phi/2\pi=1/2$).
  The arrows show the expectation values of the bond currents, and their lengths are proportional to the magnitudes.
  The middle region of $L=60$ is shown.}
  \label{fig:current_square}
\end{figure*}

\begin{figure}[h]
  \centering
  \includegraphics[width=\columnwidth]{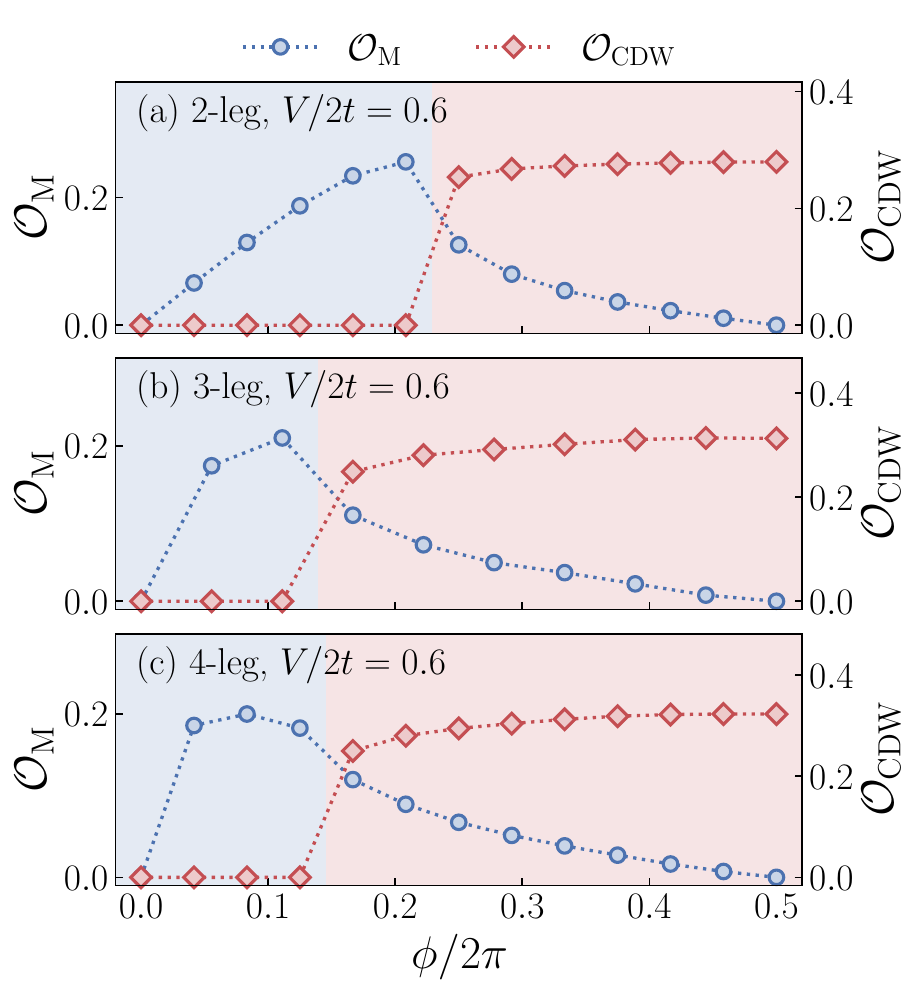}
  \caption{Magnetic flux dependence of the diamagnetic current $\mathcal{O}_{\rm M}$ (blue circles, left axis) and the CDW order parameter $\mathcal{O}_{\rm CDW}$ (red diamonds, right axis) at $V/2t=0.6$ and $L=60$ for (a) two-, (b) three-, and (c) four-leg square ladders.
  The blue and red regions indicate the Meissner and CDW phases, respectively.
  The CDW order occurs near the region where
$O_{\rm M}$ begins to decrease.}
  \label{fig:M_CDW_square}
\end{figure}

We discuss the square ladder systems in this section, and study the triangular ladders in the next section (Sec.~\ref{sec:triangular}).
To describe the magnetic flux responses of these few-leg ladders, we call a phase the Meissner-like phase when it has a finite diamagnetic leg current $\mathcal{O}_{\rm M}$ [Eq.~\eqref{eq:Jc}] and no rung current $|\mathcal{J}_{\rm rung}|$, and the vortex-like phase when it has a finite rung current $|\mathcal{J}_{\rm rung}|$ [Eq.~\eqref{eq:Jr}] with a periodic pattern, i.e., a nonzero vortex density $\rho_{\rm v}$ [Eq.~\eqref{eq:rhov}].
These phases are defined by the current patterns, and we add ``-like'' for two reasons.
First, in few-leg ladders, phases with the Meissner current and the vortex current can be either gapless superfluids or Mott insulators.
For example, the Meissner-like phase includes both the Meissner superfluid and the Meissner--Mott insulator, and the vortex-like phase includes both the vortex superfluid and the vortex--Mott insulator~\cite{Dhar2012,Dhar2013,Petrescu2013,Piraud2015,Greschner2015,Kolley2015,Natu2015}, so that they are mixed in the phase diagrams of the ladders (see Appendix~\ref{app:OYA} for details).
It is therefore difficult to discuss the Type-I and Type-II responses in a strict sense on the ladders.

Second, the Meissner-like phase on the ladders should be distinguished from the Meissner effect of bulk superfluids (superconductors) in a magnetic field.
In conventional charged superfluids (superconductors), the Meissner effect results from the perfect screening of an external magnetic field, and the length scale of this screening is the penetration depth.
In this study, the Hamiltonian~\eqref{eq:HCB} is defined with a fixed Peierls phase, which describes a magnetic flux penetrating the bulk uniformly, and this magnetic flux is a fixed external (non-dynamical) field that is not modified by the current of the bosons.
Therefore, the screening does not occur, and neither the Meissner effect in the strict sense nor the penetration depth is defined.
For convenience, we thus use the terms Type-I-like and Type-II-like, by analogy with conventional superfluids (superconductors), only as a classification of the phase sequence under increasing magnetic flux $\phi$.
We call a response Type-II-like when the Meissner-like phase is followed by a vortex-like phase, and Type-I-like when it is followed directly by a non-vortex phase.
For brevity, we simply write the Meissner phase and the vortex phase for the Meissner-like and vortex-like phases in the following.
The two-dimensional limit, where the Type-I and Type-II responses can be discussed in the strict sense, is discussed in Sec.~\ref{sec:2d}.

Figure~\ref{fig:flux_response} summarizes the magnetic flux responses of square ladders with different numbers of legs, and Fig.~\ref{fig:current_square} shows the real-space patterns of the bond currents in the four-leg ladder.
At $V=0$, increasing $\phi$ basically produces the conventional sequence from a Meissner phase to a vortex phase.
In the Meissner phase [Fig.~\ref{fig:current_square}(a)], the currents flow along the two edge legs in opposite directions, forming a chiral edge current which circulates around the ladder, while the currents on the inner legs and the rungs are small.
This edge current is the diamagnetic response to the flux, and the diamagnetic leg current $\mathcal{O}_{\rm M}$ increases monotonically with $\phi$, as in a standard superfluid.
A tiny but finite rung current can exist in the Meissner phase of a finite-size system, due to the open boundaries. This rung current is expected to vanish in the thermodynamic limit and is negligible compared with that in the vortex phase.
Beyond a critical magnetic flux, the currents penetrate the bulk [Fig.~\ref{fig:current_square}(b)]: the rung currents $|\mathcal{J}_{\rm rung}|$ grow and the inner legs carry currents, which form a periodic array of the circulating currents along the legs, i.e., vortices, and $\mathcal{O}_{\rm M}$ decreases from its maximum.
The vortex phase thus appears at nearly $V=0$ for all the numbers of legs studied, where the flux penetrates the bulk as vortices while the superfluid persists, and this behavior is a Type-II-like response.

At finite $V$, the magnetic flux response changes qualitatively.
A CDW order develops and strongly suppresses the rung current, so that the vortex phase is restricted to a region near $V=0$.
Figure~\ref{fig:M_CDW_square} shows the diamagnetic current $\mathcal{O}_{\rm M}$ and the CDW order parameter $\mathcal{O}_{\rm CDW}$ at $V/2t=0.6$.
For all $N_{\rm leg}$, $\mathcal{O}_{\rm M}$ increases monotonically at small $\phi$, reaches a maximum at $\phi_{\rm M}$, and then decreases.
The CDW order parameter $\mathcal{O}_{\rm CDW}$ vanishes in the Meissner regime and arises around the magnetic flux where $\mathcal{O}_{\rm M}$ starts to decrease.
The decrease of $\mathcal{O}_{\rm M}$ at this $V$ is therefore accompanied not by vortices but by the development of the CDW order.
To determine whether this CDW phase is non-superfluid, we also calculate the superfluid correlation function $C^{\rm SF}_r$ [Eq.~\eqref{eq:CSF}].
As shown in Appendix~\ref{app:CDW_square}, $C^{\rm SF}_r$ decays exponentially in the CDW phase for all $N_{\rm leg}$, which means that the CDW phase has no superfluid correlation and is not a supersolid, in which the superfluid order coexists with the density order.

However, we note that in few-leg ladders the magnetic flux can induce CDW order even at $V=0$, i.e., without any explicit intersite interaction.
For example, the flux-induced CDW order at $V=0$ is found in the three-leg ladder~\cite{Takayuki2026}, and similar signatures are also seen in the four-leg ladder [Figs.~\ref{fig:flux_response}(g),(k)].
This phenomenon is a nontrivial effect of quantum fluctuations, which is not simply attributed to the interaction $V$, and its signature in the four-leg ladder suggests that it is not specific to the three-leg ladder.
This does not change the Type-II-like response at nearly $V=0$, where the vortex phase still appears.

Combining these results, we conclude that the magnetic flux response evolves from Type-II-like to Type-I-like as $V$ increases.
The Meissner phase is robust against the CDW order, whereas the vortex phase is highly fragile and is suppressed by a relatively weak nearest-neighbor interaction $V$.
These trends occur in the two-, three-, and four-leg ladders (Figs.~\ref{fig:flux_response} and~\ref{fig:M_CDW_square}). The vortex region remains confined to small $V$ for all $N_{\rm leg}$ studied.
We discuss the magnetic flux response in the limit of the two-dimensional square lattice in Sec.~\ref{sec:2d}.

Besides the Type-I-like and Type-II-like classification, the square ladders with three or more legs show another flux-induced phenomenon near $\pi$-flux.
As the number of legs increases, the staggered current $|\mathcal{J}_{\rm sc}|$ [Eq.~\eqref{eq:Jsc}] appears at weak $V$ and near $\pi$-flux ($\phi/2\pi=1/2$) [Figs.~\ref{fig:flux_response}(h),(l)].
Figure~\ref{fig:current_square}(c) shows that the currents on the legs alternate in direction on every bond along the legs, and that the circulating currents alternate their chirality on neighboring plaquettes, i.e., the unit cell is doubled along the legs.
A staggered-current phase on three-leg ladders has been discussed at half filling and one-third filling~\cite{Takayuki2026,Kolley2015}.
As $V$ increases, it is again driven out by the CDW phase, but it survives at small $V$, and its region and magnitude grow from the three-leg to the four-leg ladder.
At $\pi$-flux, the Peierls phase in Eq.~\eqref{eq:gauge} is $0$ or $\pi$ on every bond, so that the Hamiltonian is real and time-reversal symmetry is preserved.
Since the current is odd under the time reversal, the finite current at $\pi$-flux means the spontaneous breaking of a time-reversal symmetry.
Moreover, as shown in the Appendix~\ref{app:chiral}, the superfluid correlation function decays algebraically, so that this phase is gapless.
Since the time-reversal symmetry is spontaneously broken while the quasi-long-range superfluid order remains, this region is understood as a chiral superfluid state.

\begin{figure*}[!t]
  \centering
  \includegraphics[width=1.0\textwidth]{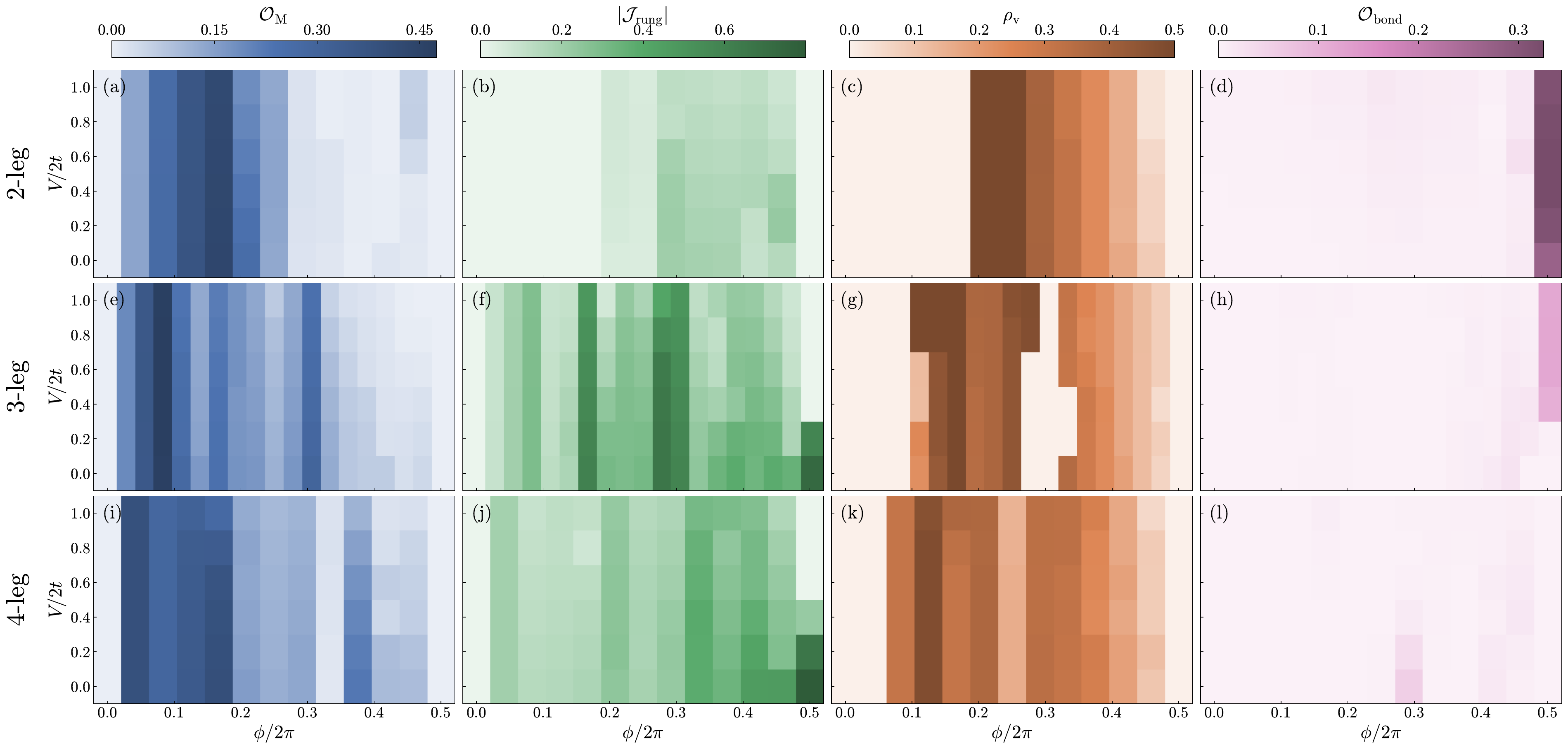}
  \caption{Magnetic flux response of (top) two-,
(middle) three-, and (bottom) four-leg
triangular ladders for $L=60$.
  From left to right, the panels show the diamagnetic current $\mathcal{O}_{\mathrm{M}}$, the rung-current magnitude $|\mathcal{J}_{\mathrm{rung}}|$, the vortex density $\rho_{\mathrm{v}}$, and the bond order parameter $\mathcal{O}_{\mathrm{bond}}$, as functions of $\phi/2\pi$ and $V/2t$.}
  \label{fig:tri_phase}
\end{figure*}

\begin{figure*}[t]
  \centering
  \includegraphics[width=\textwidth]{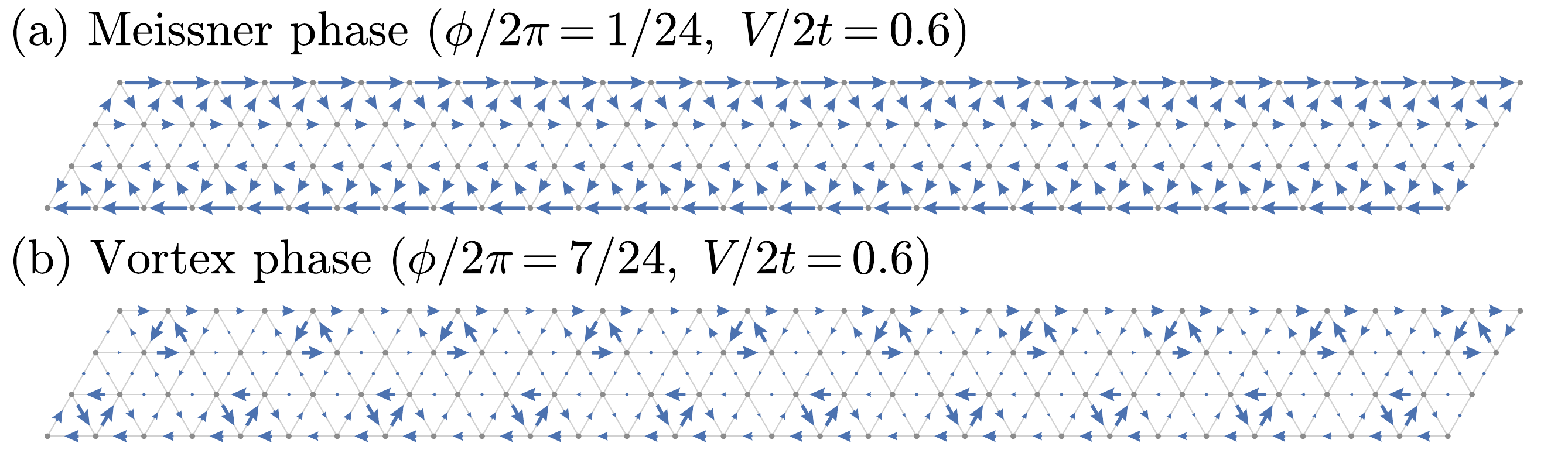}
  \caption{Real-space patterns of the bond currents in the four-leg triangular ladder at $V/2t=0.6$ for (a) the Meissner phase ($\phi/2\pi=1/24$) and (b) the vortex phase ($\phi/2\pi=7/24$).
  The arrows show the expectation values of the bond currents, including the diagonal bonds, and their lengths are proportional to the magnitudes.
  The middle region of $L=60$ is shown.}
  \label{fig:current_triangular}
\end{figure*}

\section{Magnetic flux response in triangular ladders}
\label{sec:triangular}

To clarify how the lattice geometry affects the magnetic flux response, we investigate triangular ladders, where the diagonal bonds introduce geometric frustration.
Triangular ladders under a magnetic flux have also been studied in recent years~\cite{Halati2023,Li2023,Baldelli2024,Zhang2026,Dasgupta2026,Jian2026}.
As in the square ladders, the two-leg triangular ladder is known to show the sequence from the Meissner phase to the vortex phase with increasing magnetic flux.
Here, we discuss the effects of the nearest-neighbor interaction for ladders with up to four legs.

\begin{figure}[b]
  \centering
  \includegraphics[width=\columnwidth]{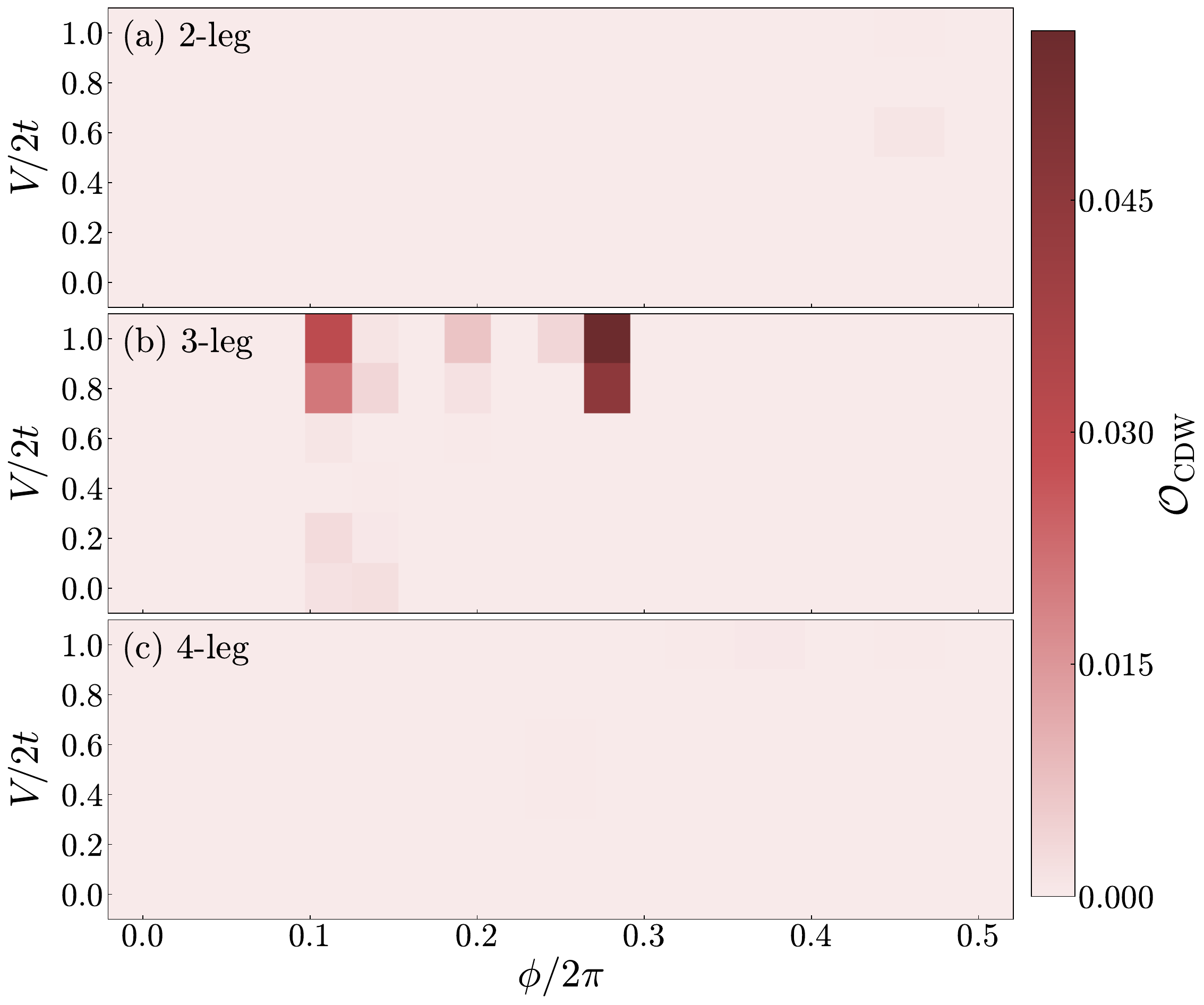}
  \caption{CDW order parameter $\mathcal{O}_{\rm CDW}$ of the triangular ladders with (a) two-, (b) three-, and (c) four-leg as a function of $\phi/2\pi$ and $V/2t$ for $L=60$.}
  \label{fig:CDW_heatmap_triangular}
\end{figure}

\begin{figure}[t]
  \centering
  \includegraphics[width=\columnwidth]{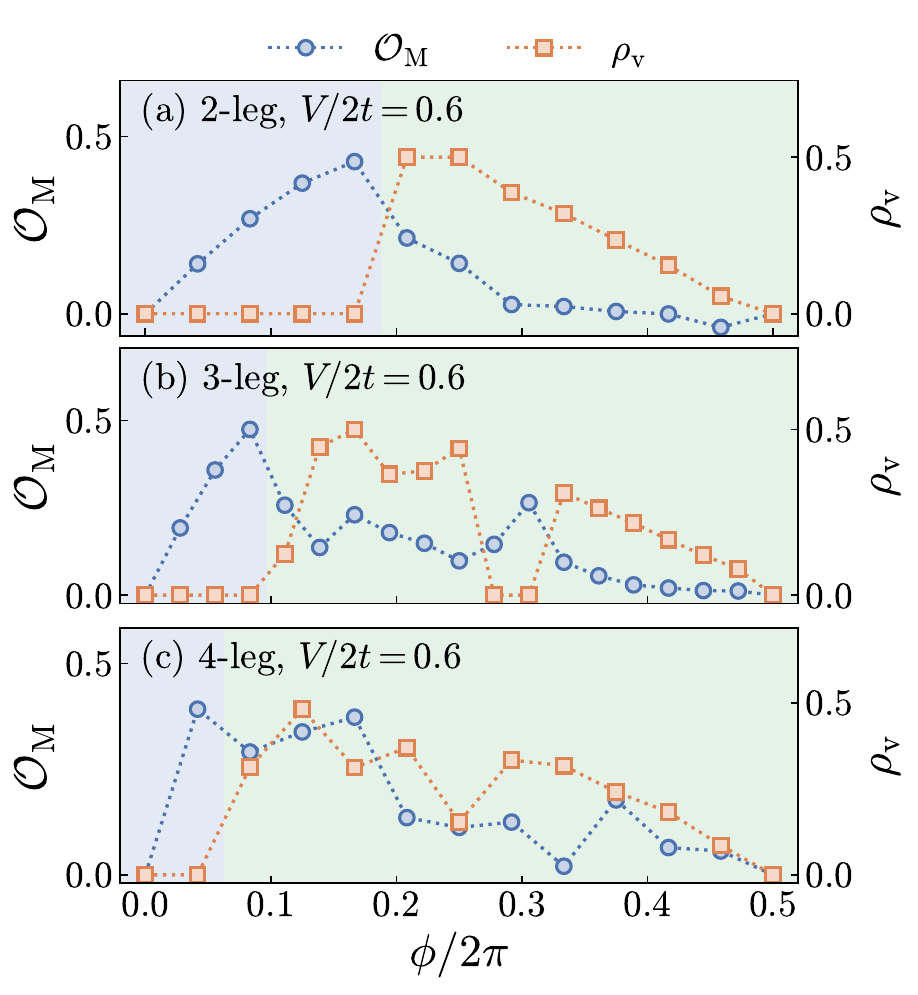}
  \caption{Magnetic flux dependence of the diamagnetic current $\mathcal{O}_{\rm M}$ (blue circles, left axis) and the vortex density $\rho_{\rm v}$ (orange squares, right axis) at $V/2t=0.6$ and $L=60$ for (a) two-, (b) three-, and (c) four-leg triangular ladders.
  The blue and green shaded regions indicate the Meissner and vortex phases, respectively.}
    \label{fig:M_vor_triangular}
\end{figure}

Figure~\ref{fig:tri_phase} summarizes the magnetic flux response of two-, three-, and four-leg triangular ladders in terms of the diamagnetic current $\mathcal{O}_{\rm M}$ [Eq.~\eqref{eq:Jc}], the rung current $|\mathcal{J}_{\rm rung}|$ [Eq.~\eqref{eq:Jr}], the vortex density $\rho_{\rm v}$ [Eq.~\eqref{eq:rhov}], and the bond order $\mathcal{O}_{\rm bond}$ [Eq.~\eqref{eq:bond}].
Figure~\ref{fig:current_triangular} shows the real-space patterns of the bond currents in the four-leg ladder: in the Meissner phase, the currents flow along the two edge legs [Fig.~\ref{fig:current_triangular}(a)], whereas in the vortex phase, the circulating currents appear periodically in the bulk [Fig.~\ref{fig:current_triangular}(b)].
The triangular ladders show a Meissner phase in which $\mathcal{O}_{\rm M}$ increases with $\phi$, followed by a non-Meissner phase in which $\mathcal{O}_{\rm M}$ decreases and the rung current is finite.
This is most clearly seen in the two-leg ladder, where the peak position of $\mathcal{O}_{\rm M}$ is nearly independent of $V$ [Fig.~\ref{fig:tri_phase}(a)], as shown at $V/2t=0.6$ in Fig.~\ref{fig:M_vor_triangular}(a), while the growth of $|\mathcal{J}_{\rm rung}|$ at larger $\phi$ signals the Meissner-to-vortex crossover [Fig.~\ref{fig:tri_phase}(b)].
As the number of legs increases, the window of the Meissner phase shrinks, although $\mathcal{O}_{\rm M}$ still increases at small $\phi$ and decreases at larger $\phi$ [Figs.~\ref{fig:tri_phase}(a),(e),(i) and~\ref{fig:M_vor_triangular}(a)-(c)].
The shrinking Meissner window is consistent with the absence of the Meissner effect in the two-dimensional limit (see Sec.~\ref{sec:2d}).

In the triangular ladders, the decrease of $\mathcal{O}_{\rm M}$ is accompanied by the growth of the rung current over a wide range of the nearest-neighbor interaction $V$ [Fig.~\ref{fig:tri_phase}].
To examine this behavior, we calculate the vortex density $\rho_{\rm v}$ [Eq.~\eqref{eq:rhov}].
As shown in Figs.~\ref{fig:tri_phase}(c),(g),(k) and Fig.~\ref{fig:M_vor_triangular}, $\rho_{\rm v}$ suddenly increases where the Meissner current $\mathcal{O}_{\rm M}$ starts to decrease even at finite $V$.
Since a periodic current flows in the bulk there, it is natural to identify this region as a vortex phase, which is consistent with the behavior of a Type-II superconductor.
In the three-leg ladder, $\rho_{\rm v}$ drops to zero at $\phi/2\pi\approx0.3$ [Fig.~\ref{fig:M_vor_triangular}(b)], but the large $|\mathcal{J}_{\rm rung}|$ [Fig.~\ref{fig:tri_phase}(f)] indicates a commensurate vortex phase with a maximally packed vortex pattern (see Sec.~\ref{sec:Observables} for details).

In contrast to the square ladders, in which increasing $V$ drives the Meissner phase directly into the CDW phase, the vortex phase on the triangular ladders is robust against the nearest-neighbor interaction $V$.
The rung current remains dominant over a wide range of $\phi$ for all $V$ studied, and the CDW order is strongly suppressed (Fig.~\ref{fig:CDW_heatmap_triangular} and Appendix~\ref{app:CDW_triangular}).
The only exception is the three-leg ladder, in which $\mathcal{O}_{\rm CDW}$ is finite at some parameters [Fig.~\ref{fig:CDW_heatmap_triangular}(b)].
This CDW order appears only at isolated values of the flux, does not replace the vortex phase over the whole range, and does not persist in wider ladders (Appendix~\ref{app:CDW_triangular_3leg}).
Therefore, the Meissner phase is not directly replaced by a non-vortex phase for any $V$ studied, and the response of the triangular ladders remains Type-II-like.

Two limitations of the present analysis should be mentioned.
First, in the decreasing region, $\mathcal{O}_{\rm M}$ is not monotonic in $\phi$ but shows irregular oscillations [Fig.~\ref{fig:M_vor_triangular}(a)-(c)].
A similar behavior was also reported in previous work on the two-leg ladder~\cite{Halati2023}, and its origin is not clear at present.
Second, we have not distinguished the commensurate and the incommensurate vortex phases~\cite{Kolley2015, Takayuki2026}.
We only claim the existence of the Meissner phase (in which $\mathcal{O}_{\rm M}$ increases with $\phi$), and of the vortex phase (in which $\mathcal{O}_{\rm M}$ decreases with $\phi$ and $\rho_{\rm v}$ appears).

Near $\pi$-flux ($\phi/2\pi=1/2$), a distinct phase appears in the two- and three-leg ladders, in which $\mathcal{O}_{\rm M}$ and $|\mathcal{J}_{\rm rung}|$ vanish and the bond order parameter $\mathcal{O}_{\rm bond}$ [Eq.~\eqref{eq:bond}] becomes finite [Figs.~\ref{fig:tri_phase}(d),(h)].
In the two-leg triangular ladder, this bond-ordered phase is already present at $V=0$~\cite{Halati2023}.
In the two-dimensional limit, the model at $\pi$-flux is the antiferromagnetic triangular XXZ model, whose ground state is known to exhibit superfluid order ($120^\circ$ Néel order in the spin-language) without bond order, so that the bond-ordered phase is expected to be a finite-width effect of the narrow ladders.
Since it is not directly related to the Type-I/II classification, the detailed analysis of the dimer structure is given in Appendix~\ref{app:dimer}.

In summary, the triangular ladders show a magnetic flux response that is qualitatively different from that of the square ladders. In the triangular ladders, the vortex phase is robust against $V$ and the CDW order is strongly suppressed, so that the response remains Type-II-like, in contrast to the Type-I-like response of the square ladders at finite $V$.

\begin{figure*}[t]
\centering
\includegraphics[width=\textwidth]{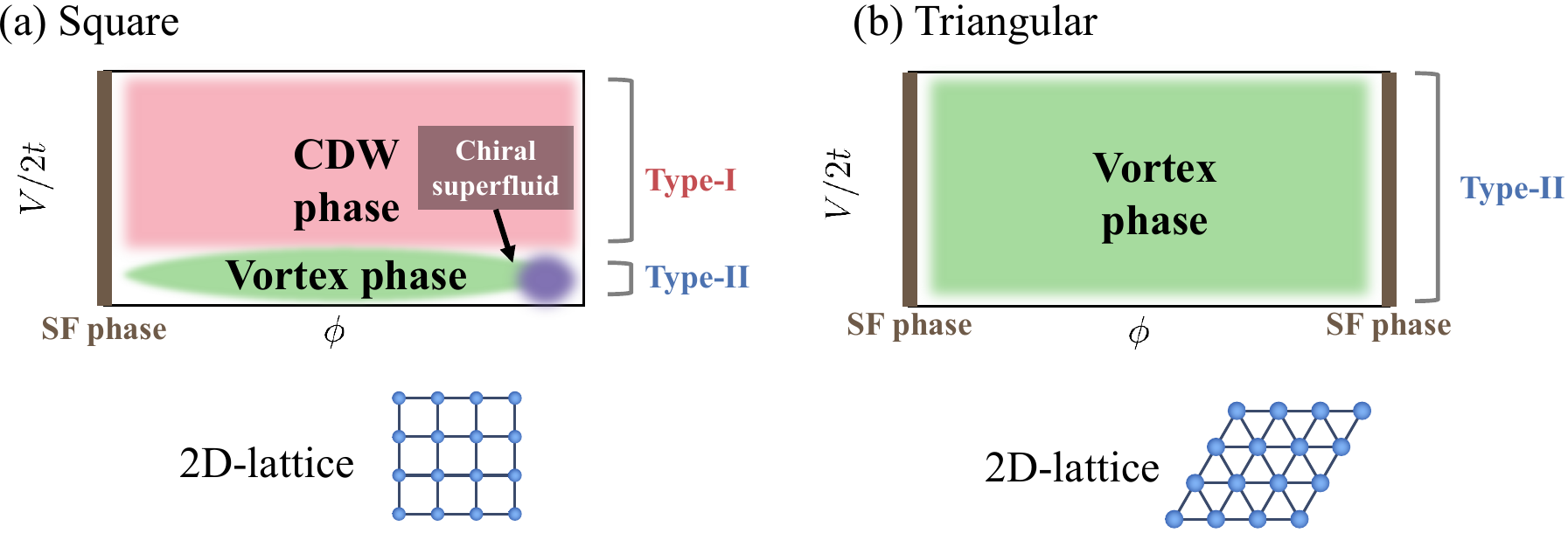}
\caption{Schematic phase diagrams and lattice geometries
for (left) square and (right) triangular lattices in
the two-dimensional limit, as conjectured from the
numerical results for few-leg ladders in the region $0\leq \phi\leq \pi$ and
$0\leq V/2t\le1$.
The square and triangular cases illustrate Type-I-like
and Type-II-like flux responses, respectively.}

\label{fig:sketch}
\end{figure*}

\section{Discussion on the two-dimensional limit}
\label{sec:2d}

Based on the ladder results in Sec.~\ref{sec:square} and Sec.~\ref{sec:triangular}, we discuss an expected magnetic flux response in two dimensions.
First, in two dimensions, the hard-core extended Bose-Hubbard model maps onto the spin-$1/2$ XXZ model, and its ground state is known in some cases at $\phi=0,\pi$.
On the square lattice, the ground state without flux ($\phi=0$) is a superfluid (XY-ordered in the spin language) for $V/2t<1$ ($J_z/J_{xy}<1$)~\cite{Kubo1998, Hebert2001, Auerbach1994}.
On the triangular lattice, the ground state without flux is also a superfluid for small $V$, and a supersolid with the three-sublattice density order does not appear for $V/2t<1$~\cite{Wessel2005,Heidarian2005,Melko2005,Boninsegni2005}.
At $\pi$-flux ($\phi/2\pi=1/2$), the triangular lattice is mapped by the gauge transformation (Sec.~\ref{sec:gauge}) to the spin model with the antiferromagnetic in-plane interaction, where there is geometric frustration.
There, the ground state is a superfluid for $V/2t<1$ and is a supersolid ($120^\circ$ N\'eel order) at the SU(2) point $V/2t=1$~\cite{Wang2009}.

We can infer the expected phase diagrams for the flux range $0\leq \phi \leq \pi$ based on the numerical results for the few-leg ladders.
Figure~\ref{fig:sketch} shows schematic phase diagrams for the two-dimensional square and triangular lattices, obtained by extrapolating the ladder results.
We suppose that the Meissner phase vanishes in the two-dimensional limit on both lattices.
Based on the ladder results, we argue that a small magnetic flux induces the CDW phase on the square lattice and the vortex phase on the triangular lattice.

For the two-dimensional square lattice, we expect a Type-I flux response at finite $V$.
A magnetic flux induces the CDW phase, and the Meissner phase turns directly into the CDW phase without the vortex phase.
Near $V=0$, the superfluid persists with vortices, which is a Type-II response.
Near $\pi$-flux ($\phi/2\pi=1/2$), the staggered current survives at weak $V$ in the three- and four-leg ladders [Figs.~\ref{fig:flux_response}(h),(l)], from which we expect a chiral superfluid in two dimensions in this region (purple region in Fig.~\ref{fig:sketch}).

The enhancement of the CDW order by the magnetic flux on the square lattice is reminiscent of the magnetic catalysis~\cite{Miransky2015}.
The magnetic catalysis is a phenomenon in which a magnetic flux enhances the effect of interactions and thereby promotes order and gap formation.
This phenomenon has been discussed mainly for Dirac fermions, such as the quark condensation in high-energy physics~\cite{Miransky2015,Gusynin1994,Gusynin1995,Krive1992,Scherer2012,Lenz2023,Fukushima2012} and quasi-relativistic electrons in honeycomb lattices~\cite{Gorbar2002,Herbut2008,Roy2011,Roy2014,Tada2020}.
It has also been examined for the superfluid state of bosons~\cite{Perez1996,Ayala2012}.
In the two-dimensional spin-$1/2$ Heisenberg model under a uniform flux, an enhancement of the N\'eel order (equivalent to CDW of bosons) similar to ours was found~\cite{Yokoyama2025}.

The triangular lattice shows a different behavior.
The vortex phase appears with the magnetic flux and extends over the whole range of $V$ studied in the ladders (Fig.~\ref{fig:tri_phase}), and the CDW order is suppressed (Appendix~\ref{app:CDW_triangular}).
We therefore expect that, in two dimensions, the superfluid phase at small $\phi$ is followed by the vortex phase irrespective of $V$, and the response is Type-II.
This robustness against $V$ may be attributed to the geometric frustration, which suppresses a simple CDW state in the presence of the nearest-neighbor repulsion.

The expected absence of magnetic flux-induced Type-I response  on the triangular lattice suggests that geometric frustration may influence the Type-I and Type-II response. 
The conventional understanding of the Type-I and Type-II superconductors is based on the ratio of the magnetic penetration depth to the coherence length.
Our results suggest that another insight, namely the interplay of the interaction and the lattice geometry, is also relevant to the emergence of the Type-I and Type-II responses in bosonic superfluids.
Superconductivity in frustrated systems such as kagome metals~\cite{Jiang2023,Wilson2024} has been extensively discussed.
Studying superfluids (superconductors) on frustrated lattices from the viewpoint of the magnetic flux response may therefore provide a new route to understanding them.

\section{Summary}
\label{sec:summary}

We studied the magnetic flux response of the hard-core extended Bose-Hubbard model at half filling on square and triangular ladders with two to four legs using DMRG.
We classified the response as Type-I-like or Type-II-like according to whether the Meissner-like phase is followed by a vortex-like phase or directly by a non-vortex phase.
On the square ladders, the vortex phase appears only in a region near $V=0$, so the response is Type-II-like only there.
At finite $V$, the Meissner phase gives way directly to a flux-induced CDW phase, which is non-superfluid, and the response becomes Type-I-like.
In contrast to the square ladders, on the triangular ladders, the vortex phase remains over a wide range of $\phi$ for all $V$ studied, the CDW order is suppressed, and the response stays Type-II-like.

Extrapolating the ladder results to two dimensions, where the Type-I and Type-II responses can be discussed in the strict sense, we propose schematic phase diagrams (Fig.~\ref{fig:sketch}).
On the square lattice, we argue that at weak $V$ the vortex phase is suppressed and instead the CDW phase is stabilized, so that the response is Type-I.
The enhancement of the CDW order by the magnetic flux is reminiscent of the magnetic catalysis.
On the triangular lattice, the geometric frustration suppresses a simple CDW state, and the vortex phase is expected to persist for any $V$, so that the response will be Type-II.
The absence of the flux-induced Type-I response on the triangular lattice suggests that, in addition to the penetration depth and the coherence length of the conventional understanding, the interplay of the interaction and the geometric frustration of the lattice may influence the Type-I and Type-II responses.

The two-dimensional phase diagrams are conjectures based on the ladder results.
Direct calculations in two dimensions, such as a larger number of legs or cylinder geometries, are needed to confirm them.
It is also an interesting future problem to clarify how the strong CDW tendency at small $V$ in the ladders connects to the two-dimensional limit.

\begin{acknowledgments}
We thank Shun Uchino for valuable discussion. The DMRG calculations were performed using ITensor.jl~\cite{ITensor}. Numerical calculations were performed using the facilities of the Supercomputer Center at the Institute for Solid State Physics, the University of Tokyo. This study is supported by JST SPRING Grant No.~JPMJSP2132, JSPS KAKENHI No.~22K03513 and JSPS KAKENHI No.~26K06946.
\end{acknowledgments}

\appendix
\section{Condition for gap opening in the ladders: the Oshikawa--Yamanaka--Affleck condition}
\label{app:OYA}

In this appendix, we discuss in which ladders a Mott insulator can appear at half filling, which supports the statement in Sec.~\ref{sec:square} that the Meissner-like and vortex-like phases can include Mott insulators.
Here, a Mott insulator means a gapped state with a unique ground state, which preserves both the U(1) symmetry and the translation symmetry.
Such a state is characterized by the charge gap
\begin{equation}
\Delta E_{\rm charge}=E(N_b+1)+E(N_b-1)-2E(N_b),
\label{eq:Egap}
\end{equation}
where $E(N_b)$ is the ground-state energy in the sector with $N_b$ bosons.

\subsection{Translation symmetry under the flux}
The Oshikawa--Yamanaka--Affleck (OYA) condition~\cite{Oshikawa1997} requires the translation symmetry along the legs.
In the gauge of Eq.~\eqref{eq:gauge}, the Peierls phase on the rung and the diagonal bonds depends on $x$, so that the simple translation $x\to x+1$ is not manifest.
However, this is an artifact of the gauge choice.
By the local U(1) gauge transformation $b_{x,y}\to e^{i\theta_{x,y}}b_{x,y}$ with $\theta_{x,y}=\kappa\,\phi\,xy$, where $\kappa=1$ for the square ladder and $\kappa=2$ for the triangular ladder, the Peierls phases are changed into
\begin{equation}
A^{\parallel}_y=-\kappa\,\phi\,y,\quad A^{\perp}_x=0,\quad A^{\times}_y=\kappa\,\phi\left(y+\tfrac{1}{2}\right),
\end{equation}
on the leg, the rung, and the diagonal bonds, respectively, which depend only on the leg index $y$ and not on $x$.
Therefore, the Hamiltonian is invariant under the translation by one site along the legs for any $\phi$, namely, the magnetic translation of the original gauge, which commutes with the total particle number.
This is because the ladder has a finite width, so that the unit cell is not enlarged by the magnetic flux, in contrast to the two-dimensional case with a rational magnetic flux.

\subsection{OYA condition}
For a spin system with the U(1) symmetry and the translation symmetry, the OYA condition~\cite{Oshikawa1997} states that a uniquely gapped ground state is possible only when
\begin{equation}
l_{\rm unit}\,N_{\rm leg}\left(S-\ev{S^z}\right)\in\mathbb{Z},
\label{eq:OYA}
\end{equation}
where $l_{\rm unit}$ is the length of the (enlarged) unit cell along the legs, $N_{\rm leg}$ is the number of spins in the unit cell with $l_{\rm unit}=1$, i.e., the number of legs, and $S$ is the spin.
When this condition is satisfied, the system can have a magnetization plateau with the translation symmetry~\cite{Cabra1997,Cabra1998}, which corresponds to a Mott insulator with a unique ground state in the boson language.
In our model, $S=1/2$ and $\ev{S^z}=\ev{n}-1/2=0$ at half filling, so that Eq.~\eqref{eq:OYA} becomes
\begin{equation}
\frac{l_{\rm unit}N_{\rm leg}}{2}\in\mathbb{Z}.
\end{equation}
This condition does not depend on $V$ or $\phi$, nor on the lattice geometry, since the diagonal bonds of the triangular ladder do not change the unit cell.
It leads to the following conclusions.

\begin{table}[t]
\caption{Whether a uniquely gapped ground state is allowed by the OYA condition at half filling, for both the square and the triangular ladders. The condition is satisfied when $l_{\rm unit}N_{\rm leg}/2$ is an integer, where $l_{\rm unit}$ is the unit cell size of the ground state.}
\label{tab:OYA}
\begin{ruledtabular}
\begin{tabular}{ccc}
$N_{\rm leg}$ & $l_{\rm unit}=1$ & $l_{\rm unit}=2$ \\
\hline
2 & allowed & allowed \\
3 & forbidden & allowed \\
4 & allowed & allowed \\
\end{tabular}
\end{ruledtabular}
\end{table}

For the two- and four-leg ladders, the condition is satisfied with $l_{\rm unit}=1$ (Table~\ref{tab:OYA}), so that a Mott insulator with a unique ground state is not excluded.
It can be combined with the Meissner current or the vortex pattern, which gives the Meissner--Mott insulator and the vortex--Mott insulator mentioned in Sec.~\ref{sec:square}.
In contrast, for the three-leg ladder, the condition is not satisfied with $l_{\rm unit}=1$, so that the ground state must be gapless or must break the translation symmetry if it is gapped.
In the latter case, the condition is satisfied with $l_{\rm unit}=2$, as in the gapped staggered CDW phase.
For all the numbers of legs, the condition is satisfied with $l_{\rm unit}=2$.
The gapped CDW phase of Appendix~\ref{app:CDW_square} is therefore consistent with the OYA condition irrespective of $N_{\rm leg}$.

The OYA condition is a necessary condition for the existence of a uniquely gapped state.
Therefore, it does not guarantee a Mott insulator in the two- and four-leg ladders, and whether it is actually realized depends on the parameters.

\begin{figure}[t]
  \centering
  \includegraphics[width=\columnwidth]{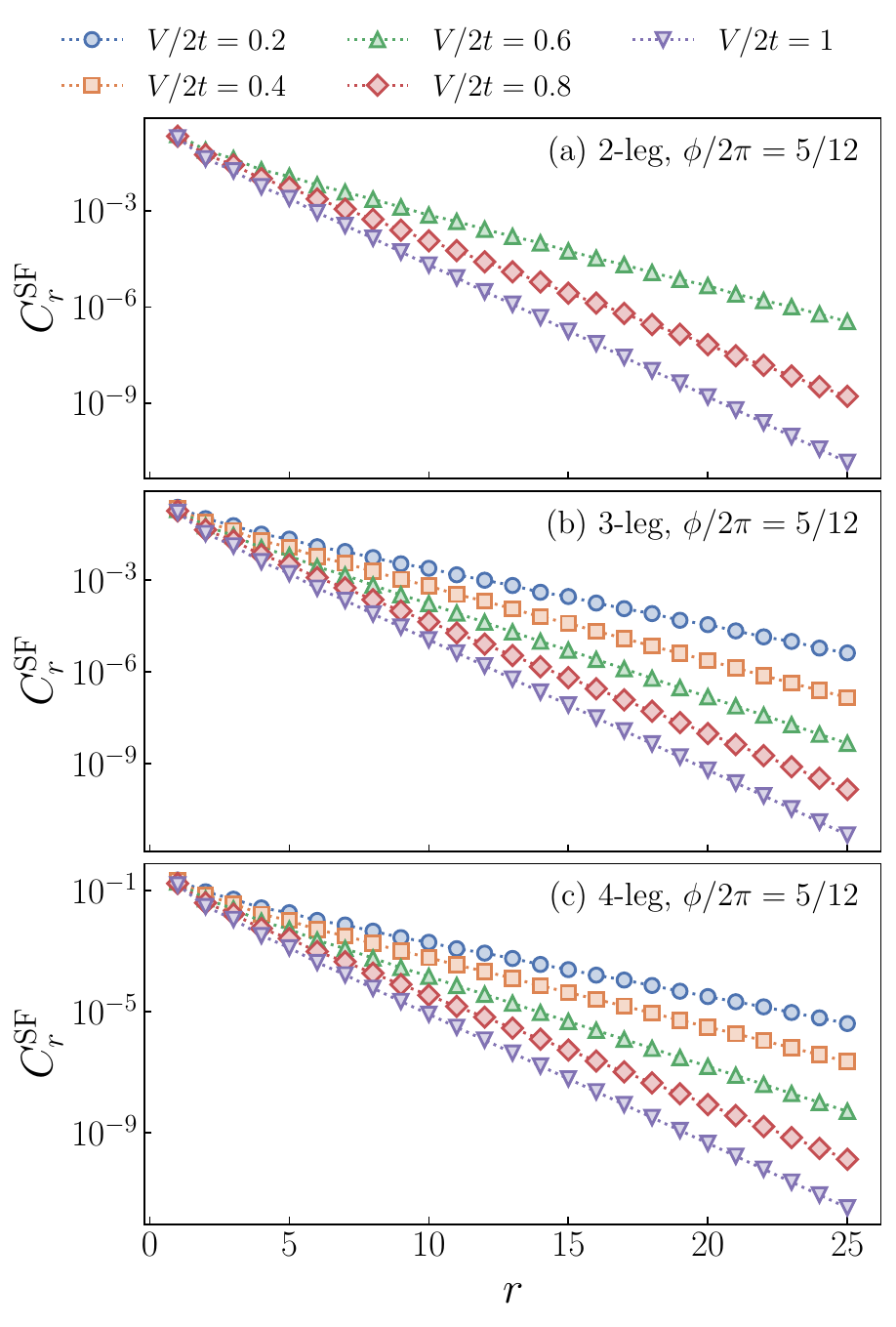}
  \caption{Superfluid correlation function $C^{\rm SF}_r$ [Eq.~\eqref{eq:CSF}] at $\phi/2\pi=5/12$ and $L=60$ for (a) two-, (b) three-, and (c) four-leg square ladders.
  The vertical axis is logarithmic, so that the straight lines indicate the exponential decay.}
  \label{fig:CDW_corr_square}
\end{figure}

\section{Absence of superfluid correlation in the CDW phase of the square ladders}
\label{app:CDW_square}

In this appendix, we show that the superfluid correlation decays exponentially in the CDW phase of the square ladders, which implies that the CDW phase is gapped and is not a supersolid, in which the superfluid order coexists with the density order.
We define the superfluid correlation function along the legs as
\begin{equation}
C^{\rm SF}_r=\frac{1}{N_{\rm leg}}\sum_{y=1}^{N_{\rm leg}}\abs{\ev{b^\dagger_{x_0,y}\,b^{}_{x_0+r,y}}},
\label{eq:CSF}
\end{equation}
where $x_0=L/2$ is the reference column.
In the gapless superfluid phase, which is a Tomonaga-Luttinger liquid in the quasi-one-dimensional ladders, $C^{\rm SF}_r$ decays algebraically, $C^{\rm SF}_r \propto r^{-\eta}$.
In a supersolid, it should also decay algebraically or approach a constant, since the superfluid quasi-long-range order coexists with the density order.
In contrast, a gapped state without the superfluid correlation shows exponential decay, $C^{\rm SF}_r \propto e^{-r/\xi}$ with a finite correlation length $\xi$.

Figure~\ref{fig:CDW_corr_square} shows the superfluid correlation function $C^{\rm SF}_r$ in the CDW phase at $\phi/2\pi=5/12 \, (=0.417)$ for the two-, three-, and four-leg ladders.
The superfluid correlation function $C^{\rm SF}_r$
decays exponentially for all $N_{\rm leg}$.
The correlation length $\xi$ appears to decrease as $V$ increases, which is consistent with the picture that the CDW order becomes stronger with the interaction and the gap becomes larger.
This result shows that the superfluid quasi-long-range order vanishes in the CDW region.
Therefore, we conclude that the CDW phase in the square ladders (Figs.~\ref{fig:flux_response} and~\ref{fig:M_CDW_square}) is not a supersolid, and the Meissner phase gives way directly to a non-superfluid phase.

\section{Superfluid with staggered current in square ladders}
\label{app:chiral}

\begin{figure}[t]
  \centering
  \includegraphics[width=\columnwidth]{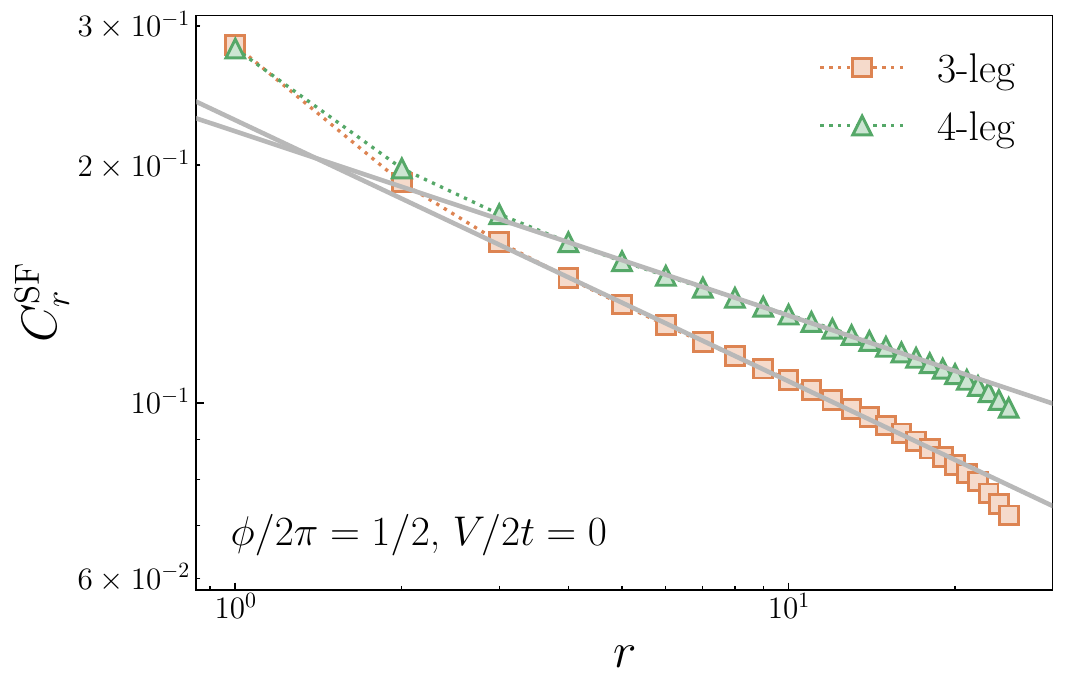}
  \caption{Superfluid correlation function $C^{\rm SF}_r$ [Eq.~\eqref{eq:CSF}] at $\phi/2\pi=1/2$ and $V/2t=0$ for the three- and four-leg square ladders with $L=60$, plotted on a log--log scale.
  The gray lines are power-law fits $C^{\rm SF}_r \propto r^{-\eta}$ in the range $5 \le r \le 20$.
  The downturn for $r \gtrsim 20$ is a boundary effect.}
  \label{fig:chiral_corr}
\end{figure}
\begin{figure*}[t]
  \centering
  \includegraphics[width=\textwidth]{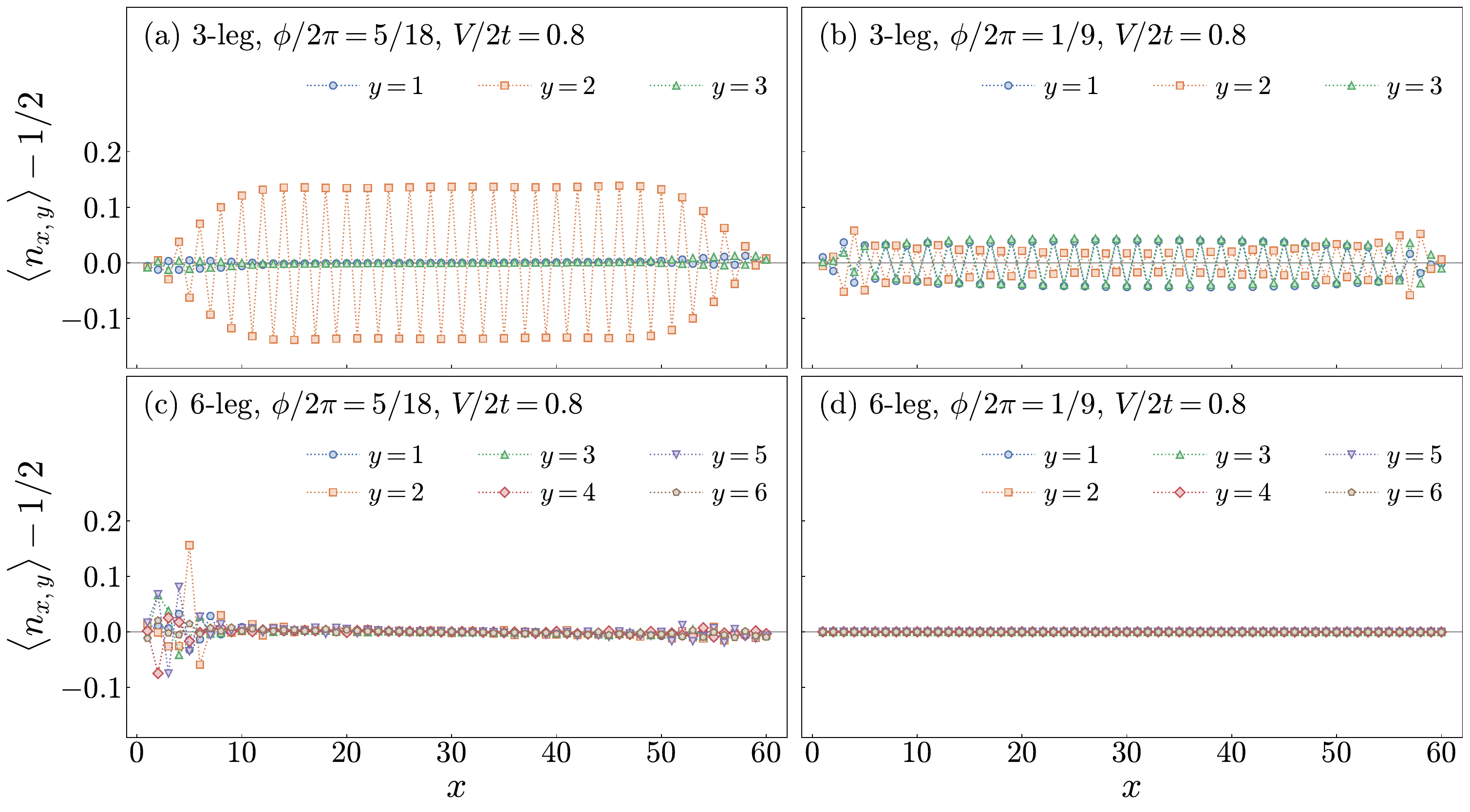}
  \caption{Density profiles $\ev{n_{x,y}}-1/2$ along the legs of the triangular ladders at $V/2t=0.8$ and $L=60$ for the three-leg ladder at (a) $\phi/2\pi=5/18$ and (b) $\phi/2\pi=1/9$, and for the six-leg ladder at (c) $\phi/2\pi=5/18$ and (d) $\phi/2\pi=1/9$.
  Different symbols show different legs $y$.}
  \label{fig:CDW_profile_triangular}
\end{figure*}

In this appendix, we show that the staggered-current phase near $\pi$-flux (Sec.~\ref{sec:square}) is a superfluid, which is required to call it a chiral superfluid.
We use the superfluid correlation function $C^{\rm SF}_r$ defined in Eq.~\eqref{eq:CSF}.
In a superfluid, which is a gapless Tomonaga-Luttinger liquid in the quasi-one-dimensional ladders, $C^{\rm SF}_r$ decays algebraically, $C^{\rm SF}_r\sim r^{-\eta}$, whereas it decays exponentially in a gapped phase such as the CDW phase (Appendix~\ref{app:CDW_square}).

Figure~\ref{fig:chiral_corr} shows $C^{\rm SF}_r$ at $\pi$-flux ($\phi/2\pi=1/2$) and $V/2t=0$ for the three- and four-leg ladders, where the staggered current $|\mathcal{J}_{\rm sc}|$ is finite [Figs.~\ref{fig:flux_response}(h),(l)].
On a log-log scale, the data follow straight lines, i.e., $C^{\rm SF}_r\sim r^{-\eta}$.
This algebraic decay indicates the quasi-long-range superfluid order, in contrast to the exponential decay in the CDW phase, and hence the staggered-current phase is gapless.
Together with the spontaneous breaking of the time-reversal symmetry at $\pi$-flux, we regard this state as a chiral superfluid.

\section{Presence and absence of the CDW order in the triangular ladders}
\label{app:CDW_triangular}

In the triangular ladders, the CDW order parameter is almost zero in the entire parameter region except for the three-leg ladder [Fig.~\ref{fig:CDW_heatmap_triangular}].
In this appendix, we first describe the CDW order which appears in the three-leg ladder, and then show that the CDW correlation is not dominant when the superfluid correlation exists, and that the phases of the triangular ladders are not supersolids.

\subsection{CDW order in the three-leg ladder}
\label{app:CDW_triangular_3leg}

Figure~\ref{fig:CDW_heatmap_triangular}(b) shows that, in the three-leg ladder, $\mathcal{O}_{\rm CDW}$ is finite at $\phi/2\pi\approx0.1$, $0.19$, and $0.28$, mainly at $V/2t\gtrsim0.7$, and that it is largest at $\phi/2\pi=5/18\ (\approx0.28)$.
At other flux values, it does not appear even in the three-leg ladder.
To investigate this CDW phase, we also calculate the ground state in the six-leg triangular ladder.
Figure~\ref{fig:CDW_profile_triangular} shows the density profiles $\ev{n_{x,y}}-1/2$ at $V/2t=0.8$, which is inside this region.
At $\phi/2\pi=5/18$ [Fig.~\ref{fig:CDW_profile_triangular}(a)], the CDW order is localized in the middle leg ($y=2$).
At $\phi/2\pi=1/9$ [Fig.~\ref{fig:CDW_profile_triangular}(b)], the modulation appears in all three legs, which is a checkerboard-type pattern as in the square ladders.
The amplitude is much smaller than that at $\phi/2\pi=5/18$.

In contrast, the CDW order disappears at both parameters in the six-leg ladder.
At $\phi/2\pi=5/18$ [Fig.~\ref{fig:CDW_profile_triangular}(c)], a modulation exists only near the left edge ($x\lesssim 10$) and decays quickly into the bulk, indicating an edge effect due to the open boundary conditions.
At $\phi/2\pi=1/9$ [Fig.~\ref{fig:CDW_profile_triangular}(d)], the density is uniform within numerical accuracy.
Since the CDW order is absent in the two-, four-, and six-leg ladders, and exists only in the three-leg ladder, it is a finite-width effect specific to the three-leg ladder.
We therefore expect that the CDW order is suppressed in the two-dimensional limit of the triangular lattice at $V/2t < 1.0$.

\subsection{Negligible CDW correlation in the superfluid phases}
\label{app:CDW_triangular_corr}

\begin{figure}[t]
  \centering
  \includegraphics[width=\columnwidth]{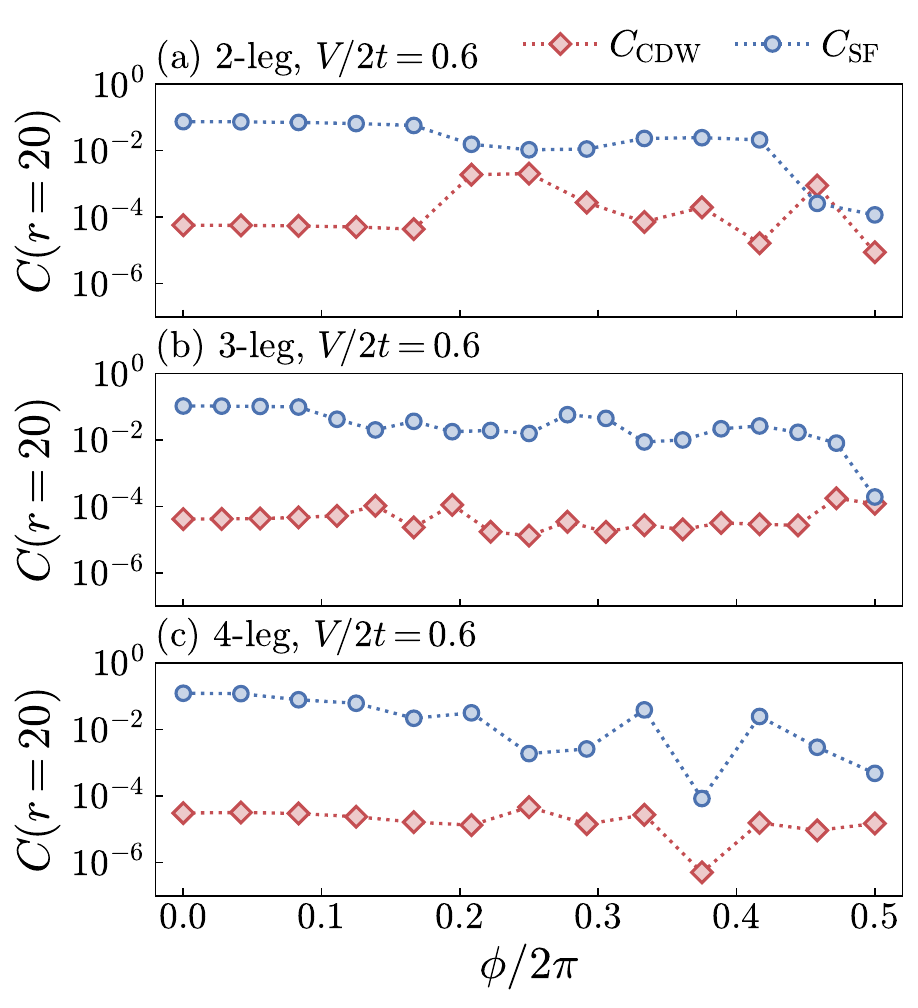}
  \caption{CDW correlation function $C^{\rm CDW}_r$ (red diamonds) and superfluid correlation function $C^{\rm SF}_r$ (blue circles) at $r=20$ as functions of $\phi/2\pi$ at $V/2t=0.6$ and $L=60$ for (a) two-, (b) three-, and (c) four-leg triangular ladders.}
  \label{fig:SF_CDW_triangular}
\end{figure}

Next, we compare the CDW correlation with the superfluid correlation.
We define the CDW correlation function along the legs as
\begin{equation}
C^{\rm CDW}_r=\frac{1}{N_{\rm leg}}\sum_{y=1}^{N_{\rm leg}}\abs{\ev{\left(n_{x_0,y}-\tfrac12\right)\left(n_{x_0+r,y}-\tfrac12\right)}},
\label{eq:CCDW}
\end{equation}
where $x_0=L/2$ is the reference column as in Eq.~\eqref{eq:CSF}.

Figure~\ref{fig:SF_CDW_triangular} shows $C^{\rm CDW}_r$ and $C^{\rm SF}_r$ at $r=20$ as functions of $\phi$ at $V/2t=0.6$.
In the regions where the superfluid correlation is sizable, $C^{\rm SF}_r\gtrsim10^{-2}$, the CDW correlation is $C^{\rm CDW}_r\lesssim2\times10^{-3}$, which is smaller than $C^{\rm SF}_r$ by at least a factor of five and in most cases by two to four orders of magnitude.
If the system exhibited long-range CDW order with the staggered amplitude $m$, $C^{\rm CDW}_r$ would approach $m^2$ at large $r$.
In the CDW phase of the square ladders, $m\approx\mathcal{O}_{\rm CDW}\approx0.3$ (Fig.~\ref{fig:M_CDW_square}), which corresponds to $m^2\approx0.09$.
The values $C^{\rm CDW}_r\lesssim2\times10^{-3}$ in the triangular ladders are smaller than this by more than one order of magnitude, and in most cases $C^{\rm CDW}_r\sim10^{-5}$--$10^{-4}$ is smaller by three orders of magnitude.
Thus, the density order is not developed where the superfluid correlation exists, so that the vortex phases of the triangular ladders are superfluids without the density order, and not supersolids in which the superfluid order coexists with the density order.

The superfluid correlation decreases near $\pi$-flux in the two- and three-leg ladders [Figs.~\ref{fig:SF_CDW_triangular}(a),(b)].
This decrease is due to the gapped dimer phase discussed in Sec.~\ref{sec:triangular}, and the CDW correlation does not increase accordingly, so that it does not indicate a CDW state.
In the two-leg ladder, $C^{\rm CDW}_r$ exceeds $C^{\rm SF}_r$ at $\phi/2\pi\approx0.46$, but both are small ($\lesssim10^{-3}$) in the crossover to the dimer phase, and thus this is not a CDW order either.

\section{Bond order in the triangular ladders at $\pi$-flux}
\label{app:dimer}

\begin{figure*}[t]
  \centering
  \includegraphics[width=\textwidth]{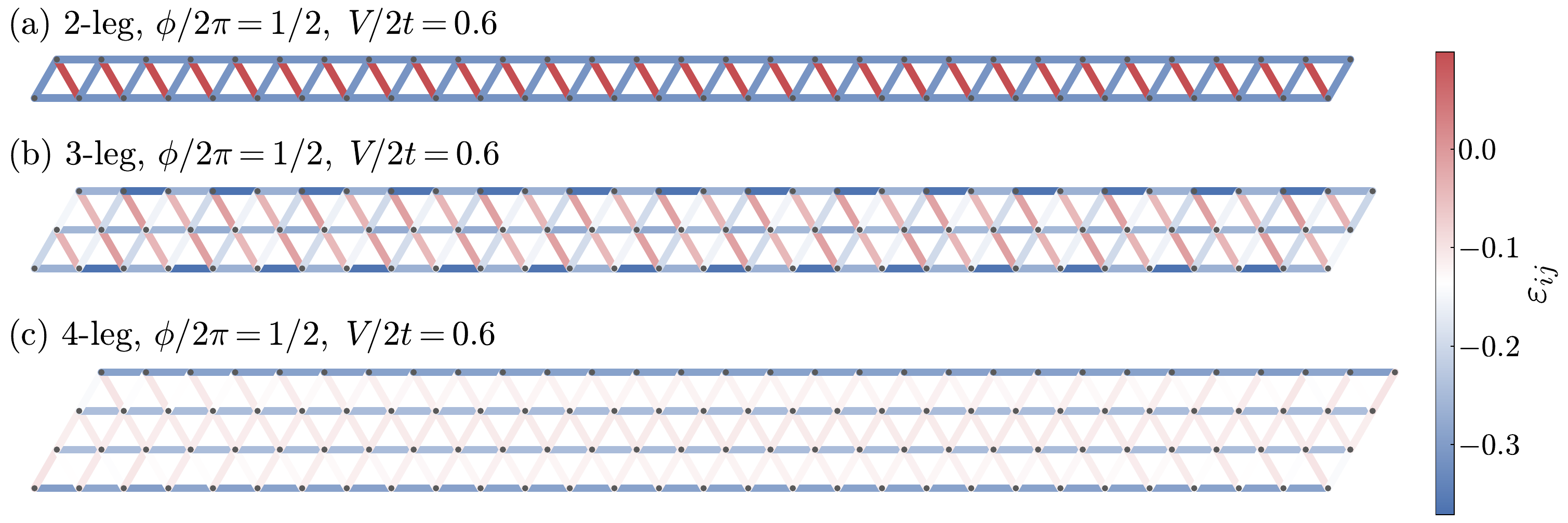}
  \caption{Bond energies of the (a) two-, (b) three-, and (c) four-leg triangular ladders at $\pi$-flux ($\phi/2\pi=1/2$) and $V/2t=0.6$.
  The color of the bonds shows the bond energy $\epsilon_{ij}$, and the negative (blue) and positive (red) values indicate the energetically favorable and unfavorable bonds, respectively.}
  \label{fig:bond_energy_pi}
\end{figure*}

In this appendix, we describe the structure of the bond energies at $\pi$-flux ($\phi/2\pi=1/2$), where the bond order appears in the two- and three-leg triangular ladders (Sec.~\ref{sec:triangular}).
At $\pi$-flux, the model is equivalent to the spin model with the antiferromagnetic in-plane interaction (Sec.~\ref{sec:gauge}).
In this form, the triangular ladder looks like an array of $N_{\rm leg}-1$ frustrated $J_1$--$J_2$ zigzag systems, in which the rung and the diagonal bonds between the neighboring legs form the nearest-neighbor bonds ($J_1$) of a chain and the leg bonds are the next-nearest-neighbor bonds ($J_2$), and the neighboring chains share the legs.
Therefore, the orders known in the frustrated zigzag systems are expected to appear readily.
Figure~\ref{fig:bond_energy_pi} shows the bond energies $\varepsilon_{ij}$ at $\phi/2\pi=1/2$ ($\pi$-flux) and $V/2t=0.6$.

The bond order at $\pi$-flux appears in the two- and three-leg ladders in different forms.
In the two-leg ladder [Fig.~\ref{fig:bond_energy_pi}(a)], the rung bonds and the leg bonds have strongly negative energies, whereas the diagonal bonds have positive energies, uniformly along the legs.
Along the zigzag chain, the strong rung bonds and the weak diagonal bonds alternate, which is the dimerization of the Majumdar--Ghosh type~\cite{Majumdar1969,Sato2011}.
This state was already found at $V=0$ in the previous work~\cite{Halati2023}, and our result shows that it persists at finite $V$.
In the three-leg ladder [Fig.~\ref{fig:bond_energy_pi}(b)], the dimerization of the Majumdar--Ghosh type coexists with the bond order of the legs: in the outer legs, the strong and the weak leg bonds alternate with the period of two lattice spacings, and the diagonal bonds also alternate in magnitude with the same period.
This state breaks the translation symmetry along the legs, which is consistent with the OYA condition (Appendix~\ref{app:OYA}), according to which a gapped ground state of the three-leg ladder at half filling must break the translation symmetry.

In the four-leg ladder [Fig.~\ref{fig:bond_energy_pi}(c)], the bond energies are uniform along the legs, and no breaking of the translation symmetry is found.
Because of the antiferromagnetic interaction, the spins on the legs $y=1$ and $3$ and those on the legs $y=2$ and $4$ are arranged in a triplet-like manner, so that they effectively combine into $S=1$ degrees of freedom, and the system looks like an $S=1$ zigzag chain.
The $S=1$ zigzag chain is known to have the Haldane phase~\cite{Haldane1983,Chepiga2016}, which is a symmetry-protected topological phase~\cite{Pollmann2010, Pollmann2012}, and the observed uniform spatial profile is consistent with this possibility.
For the four-leg ladder, the OYA condition allows a uniquely gapped state with the translation symmetry (Appendix~\ref{app:OYA}), which is consistent with such a state.

In the two-dimensional limit, the ground state at $\pi$-flux is known.
The model is equivalent to the antiferromagnetic triangular XXZ model, whose ground state is a superfluid (XY-ordered, i.e., the $120^\circ$ order in the spin language) for $V/2t<1$, without dimerization (Sec.~\ref{sec:2d}).
We therefore expect that the phenomena in few-leg ladders cross over to the two-dimensional ground state as the number of legs increases.

\bibliography{ref} 
\end{document}